\documentclass[10pt, a4paper, onecolumn, gr]{google}

\usepackage{longtable}
\usepackage{array}
\usepackage{placeins}
\usepackage{fancyvrb}
\usepackage[numbers,sort&compress]{natbib}
\setlist[itemize]{leftmargin=1.3em,itemsep=2pt,topsep=3pt}
\setlist[enumerate]{leftmargin=1.5em,itemsep=2pt,topsep=3pt}
\newcommand{\NumBenchTotal}{97}
\newcommand{\NumTasksTotal}{100}
\newcommand{\NumReplicates}{3}
\newcommand{\NumRolloutsTotal}{300}

\newcommand{\NumRolloutsVerified}{295}
\newcommand{\NumBenchComplete}{97}
\newcommand{\NumBenchAllThreeReps}{92}

\newcommand{\NumFixedW}{94}
\newcommand{\NumFixedWElec}{91}
\newcommand{\NumFixedWElecReach}{78}
\newcommand{\PctFixedWElecReach}{85.7\%}

\newcommand{\MeanFixedWDDE}{-2.13}
\newcommand{\MedianFixedWDDE}{-1.35}

\newcommand{\PctFixedWSMDReach}{96.8\%}

\newcommand{\PctAtoASMDReach}{96.9\%}
\newcommand{\PctAtoASMDReachPassOne}{92.1\%}

\newcommand{\MeanAtoASMD}{-4.76}

\newcommand{\MedianInterRepSDSMD}{0.72}

\newcommand{\MedianGenTokensK}{124\text{k}}

\newcommand{\MedianWallHours}{11.3}

\newcommand{\MedianDFTCalls}{4}

\newcommand{\MedianVMCostUSD}{\$3.82}

\newcommand{\MedianLLMCostUSD}{\$6.10}
\newcommand{\MeanLLMCostUSD}{\$6.97}

\newcommand{\NumAtoAPBFReach}{95}
\newcommand{\PctAtoAPBFReach}{97.9\%}
\newcommand{\PctAtoAPBFReachPassOne}{91.4\%}

\newcommand{\MeanAtoAPBF}{-3.81}
\newcommand{\MedianAtoAPBF}{-3.24}
\newcommand{\NumFixedWPBFReach}{92}
\newcommand{\PctFixedWPBFReach}{97.9\%}
\newcommand{\MeanFixedWPBF}{-2.18}

\uselogo{}

\title{Agentic Construction of Aqueous Transition States}

\correspondingauthor{Max Eissler (\href{mailto:meissler@google.com}{meissler@google.com}), Michael P.\ Brenner (\href{mailto:mbrenner@google.com}{mbrenner@google.com})}

\author[1,2,3]{Max Eissler}
\author[2,3]{Stefan Gugler}
\author[1,4]{Julia H.\ Yang}
\author[1,2,3,5,6]{Klaus-Robert M\"uller}
\author[1,7]{Michael P.\ Brenner}

\affil[1]{Google Research}
\affil[2]{BIFOLD -- Berlin Institute for the Foundations of Learning and Data}
\affil[3]{Machine Learning Group, Technische Universit\"at Berlin}
\affil[4]{School of Chemical and Biomolecular Engineering, Georgia Institute of Technology}
\affil[5]{Department of Artificial Intelligence, Korea University}
\affil[6]{Max Planck Institute for Informatics}
\affil[7]{School of Engineering and Applied Sciences, Harvard~University}

\begin{abstract}
\boldmath
Predicting how a liquid-phase reaction mixture equilibrates using quantum chemistry simulations alone requires computing extremely accurate free-energy barriers ($\Delta G^\ddagger$) across every competing elementary step, since a $1.36\ \mathrm{kcal\ mol^{-1}}$ error in $\Delta G^\ddagger$ causes a tenfold change in reaction rates at room temperature.
Existing packages for transition-state (TS) search scale to thousands of steps but often substantially overestimate aqueous barriers, due to missing either explicit solvent-assisted pathways or low-energy conformers. 
In contrast, hand-built density functional theory (DFT) TSs can achieve $1\text{--}2\ \mathrm{kcal\ mol^{-1}}$ fidelity with a human specialist. 
In this work we propose using LLM agents to construct TSs in water at the level of human experts.
We task Gemini 4 Carbon agents to find TSs for aqueous formose reactions explored by Kua et al.\ (2013) and Kua \& Tripoli (2024) from reactant, product, and catalyst descriptions alone.
The agents propose TS geometries working in an isolated sandbox equipped with standard computational chemistry packages. In spirit of recent agentic coding and math approaches we use an independent first-principles reaction-path verifier.
Across 97 reactions, under the specialists' DFT and solvation protocol,
    a single agent run matches or lowers expert-constructed barriers within $+2.0\ \mathrm{kcal\ mol^{-1}}$ on \PctAtoAPBFReachPassOne{} of reactions (\PctAtoAPBFReach{} across best of three runs) at a median token cost of \MedianLLMCostUSD{} per run, with best-of-three barriers more than $2\ \mathrm{kcal\ mol^{-1}}$ below the published ones on 70 of 97 reactions (by up to $16\ \mathrm{kcal\ mol^{-1}}$).
Automating specialist-level 3D transition-state construction removes a key manual bottleneck toward scalable discovery of reaction mechanisms in condensed phases.
\end{abstract}

\begin{document}
\maketitle

\section{Introduction}
\label{sec:intro}

Reaction rates govern which chemical transformations occur and which products form: for instance, whether a catalyst turns over at room temperature \cite{Kozuch2011EnergeticSpan}, which stereoisomer dominates in synthesis \cite{Seeman1983CurtinHammett,Cheong2011Organocatalysis}, or how rapidly a drug candidate hydrolyzes \cite{Rautio2008Prodrugs,Waterman2005AcceleratedAging}.
Elementary reaction rates scale exponentially with the free-energy barrier $\Delta G^\ddagger$ as $\exp(-\Delta G^\ddagger / RT)$ \cite{Eyring1935,Truhlar1996TST}, with ideal gas constant $R$ and temperature $T$, spanning milliseconds at $\Delta G^\ddagger = 15\ \mathrm{kcal\ mol^{-1}}$ to decades at $30\ \mathrm{kcal\ mol^{-1}}$ at room temperature $T=298\ \mathrm{K}$. 
Because a $1.36\ \mathrm{kcal\ mol^{-1}}$ error already shifts rates tenfold, useful rate and selectivity predictions require $\Delta G^\ddagger$ within $\sim 1\text{--}2\ \mathrm{kcal\ mol^{-1}}$ \cite{Harvey2019HomogeneousScope}.
Computing $\Delta G^\ddagger$ from first principles requires locating the 3D transition state (TS) connecting reactants and products along the reaction path \cite{Eyring1935,Unsleber2020CRNReview}.
In solution, and especially in water and other protic solvents that directly participate in the reaction, finding accurate 3D TSs has so far faced a trade-off between scale and accuracy.
Automated TS-search packages \cite{Maeda2013GRRMAFIR,Zhao2021YARP,Young2021autodE,Unsleber2022Chemoton2,Gugler2026ReactionAtlas} scale to hundreds or thousands of candidate steps. 
Several of them locate and verify multi-molecular TSs using the Intrinsic Reaction Coordinates (IRCs). Explicit water has entered such workflows as a single templated proton shuttle \cite{Zhao2022YARP2,Zhao2023HCNNetwork,Zhao2023GlucosePyrolysis,Sathyaseelan2026rTCA}, as a per-family placement heuristic for hydrolysis \cite{Fahoum2026ARC}, or as waters written into the input SMILES \cite{Stuyver2024TStools}.
In all of these the number and role of explicit waters is fixed in advance; without participating solvent molecules and conformer sampling, barriers in solution are overestimated by tens of $\mathrm{kcal\ mol^{-1}}$ when proton relays are omitted \cite{Kua2013Formose,Gugler2026ReactionAtlas}.
At the same time, hybrid quantum mechanics/molecular mechanics workflows that extract active solvent clusters from explicit solvent droplets have so far been limited to approximate barriers on isolated reactions \cite{Turtscher2025Microsolvation}. 
In contrast, human specialists reach $\sim 1\text{--}2\ \mathrm{kcal\ mol^{-1}}$ accuracy \cite{Harvey2019HomogeneousScope}, but manually placing explicit solvent molecules, sampling tens of starting conformers, and diagnosing failed DFT TS searches is tedious and time-consuming \cite{Harvey2019HomogeneousScope,Young2021autodE,Madhavan2026TSAgent}, practically restricting specialist-grade liquid-phase studies to tens of reactions \cite{Kua2013Formose,Kua2024CoreFormose}. 

\begin{figure*}[!t]
  \centering
  \includegraphics[width=0.96\textwidth]{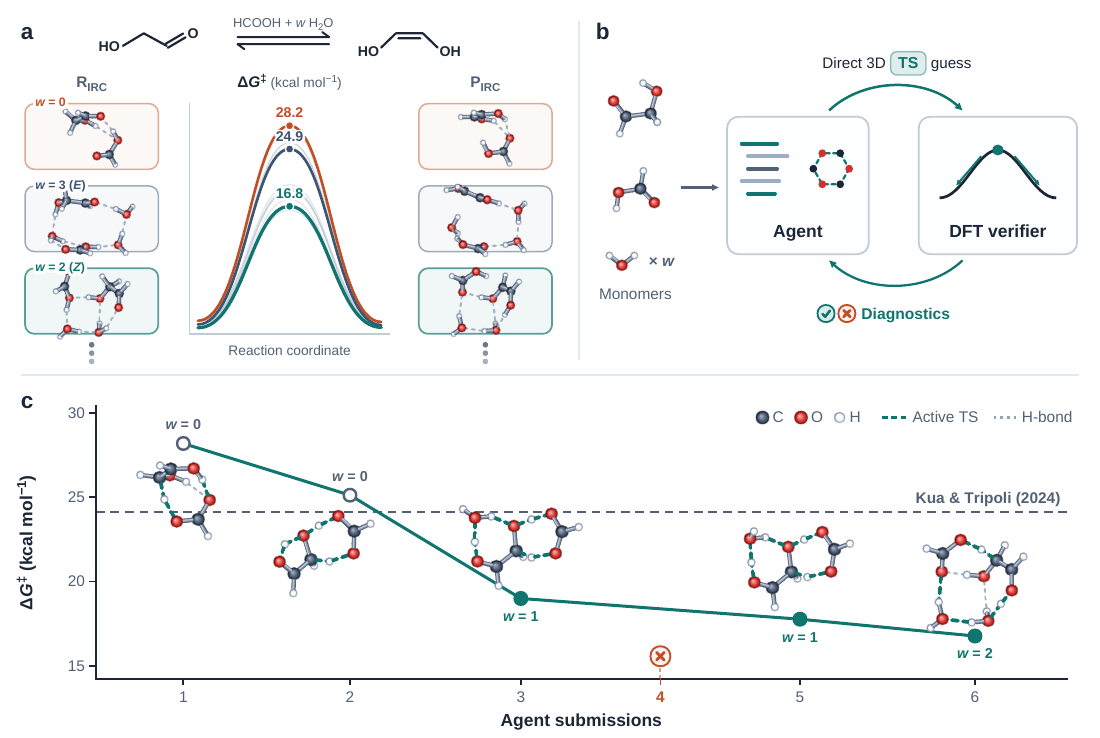}
  \caption{
  \textbf{
  Autonomous construction and first-principles verification of micro-solvated aqueous TSs.}
  \textbf{a},
  Even for a fixed reaction and catalyst (glycolaldehyde $\rightleftharpoons$ 1,2-ethenediol, catalyzed by formic acid, R42), many distinct H-bonded reactant and product intrinsic reaction coordinate ($\mathrm{IRC}$) complexes ($\mathrm{R_{IRC}} \rightleftharpoons \mathrm{P_{IRC}}$) exist across active water counts, $w$, catalyst--water relay topologies, proton-transfer atom mappings, and conformers (top to bottom: 
    $w=0$ monocyclic bridge; 
    $w=3$ unaligned \textit{E}-pathway; 
    $w=2$ pre-organized \textit{Z}-relay). 
  Because neither $w$ nor matched $(\mathrm{R_{IRC}}, \mathrm{P_{IRC}})$ endpoints are known beforehand, double-ended interpolation (e.g., NEB) cannot be initialized directly from monomers.
  \textbf{b}, 
  Given only monomer SMILES and coordinates (with any specified catalyst and $w \in \{0,\dots,5\}$ $\mathrm{H_2O}$), an offline AI agent writes Python scripts to sample $w$, relay topologies, and conformers, proposing direct 3D TS guesses without pre-optimizing encounter complexes. 
  An external B3LYP/6-311G** verifier returns diagnostics ($\checkmark$/$\times$) on saddle order, two-way $\mathrm{IRC}$ connectivity, and $\Delta G^\ddagger$.
  \textbf{c}, 
  Single-run trajectory on R42 (replicate~2; $\Delta G_{298,\mathrm{SMD}}^\ddagger$) versus Kua \& Tripoli's (2024) barrier ($24.1\ \mathrm{kcal\ mol^{-1}}$, dashed): 
  monocyclic $w = 0$ bridge (\textbf{1}, $28.2$) 
      $\rightarrow$ 
  bicyclic $w = 0$ topology (\textbf{2}, $25.1$)
      $\rightarrow$
  active $w = 1$ relay (\textbf{3}, $19.0$)
      $\rightarrow$
  rejected non-stationary $w = 2$ guess (\textbf{4}, $\times$) 
      $\rightarrow$
  refined $w = 1$ conformer (\textbf{5}, $17.8$)
      $\rightarrow$
  12-membered $w = 2$ relay (\textbf{6}, $16.8$).
  }
  \label{fig:overview}
\end{figure*}

This active solvent and co-catalyst participation creates a combinatorial search problem: a single chemical equation can be realized by many distinct 3D reactant and product encounter complexes with widely varying barriers \cite{Kua2013Formose} (Figure~\ref{fig:overview}a).
Since the number and orientation of participating molecules are unknown beforehand, double-ended interpolation (such as nudged elastic band, NEB \cite{Henkelman2000CINEB,Henkelman2000Tangent} or growing string method, GSM \cite{Zimmerman2013GSM,Zimmerman2013GSMTS}) and 3D generative TS models \cite{Duan2023OAReactDiff,Duan2025ReactOT,Kim2024TSDiffusion} do not apply directly: all require a fixed atom stoichiometry, and the double-ended methods and React-OT \cite{Duan2025ReactOT} additionally require pre-aligned reactant/product complexes.
Recent LLM-based chemistry agents \cite{Zou2025ElAgente,PerezSanchez2026ElAgenteQuntur} face the same limitation.
General quantum-chemistry agents automate calculation setup and property analysis \cite{Zou2025ElAgente,PerezSanchez2026ElAgenteQuntur}, but still need human-aligned 3D endpoints for complex TS searches \cite{PerezSanchez2026ElAgenteQuntur}.
Specialized TS-search agents also start from supplied reactant and product 3D coordinates, either on solvent-free catalyst surfaces in DFT \cite{Madhavan2026TSAgent} or in the gas phase on machine-learned potentials \cite{Meissner2026AgenticTS}.
In water, an agent must instead assemble the reactive complex from isolated monomers, choosing the active water count, $w$, proton-relay topology, and low-energy TS conformer.
Up until now, it has been unclear whether AI-based systems could construct liquid-phase TSs at human-specialist accuracy across diverse reaction classes and catalysts.

Here we show that agent-constructed TSs reach a quality comparable to a human specialist which can, however, be scaled arbitrarily across reactions. 
We use the aqueous prebiotic formose reaction set \cite{Robinson2022FormoseSelfOrg,Baltussen2024FormoseReservoir} mapped manually by Kua et al.\ (2013) \cite{Kua2013Formose} and Kua \& Tripoli (2024) \cite{Kua2024CoreFormose} as a demonstration.
Our analysis spans $46$ species and \NumBenchTotal{} unique TSs covering
    carbonyl hydration,
    aldose and ketose enolization,
    C--C aldol and retro-aldol coupling,
    Cannizzaro and Tishchenko hydride shifts,
    polyoxymethylene (POM) oligomer and furanose cyclizations, and
    catalysis by eight prebiotic co-catalysts.
Much as agent systems in mathematics and coding combine open-ended candidate generation with an independent formal verifier \cite{Hubert2025AlphaProof,RomeraParedes2024FunSearch,Aygun2026ERA}, our approach couples unconstrained Python-based TS construction in an offline sandbox with an external first-principles DFT verifier that checks saddle-point order, two-way reaction-path connectivity, and active bond participation of every explicit water and catalyst molecule (Figure~\ref{fig:overview}b).
Using diagnostic feedback from this verifier, agents iteratively refine stoichiometry and conformers from isolated monomers (Figure~\ref{fig:overview}c), matching or lowering specialist barriers within $+2.0\ \mathrm{kcal\ mol^{-1}}$ on \PctAtoAPBFReach{} of reactions across three runs (\PctAtoAPBFReachPassOne{} in a single run) under Kua's Jaguar PBF protocol at a median token cost of \MedianLLMCostUSD{} per run, while lowering barriers by more than $2\ \mathrm{kcal\ mol^{-1}}$ on 70 of 97 reactions.




Computational chemistry methods are commonly used by chemists to elucidate and explain wet lab results. 
These tasks currently require human experts to hypothesize and iteratively refine reaction mechanisms -- a labor intensive process.
We find that agents are able to perform such tasks cheaply and autonomously.

\section{Method}
\label{sec:methods}

We represent the \NumBenchTotal{} published reactions (indexed R1--R97 in Table~\ref{tab:si_master}, Appendix~\ref{app:master_table}) as \NumTasksTotal{} elementary tasks, splitting three two-step acyl condensations into separate addition and elimination tasks. We run \NumReplicates{} independent Gemini 4 Carbon agent runs (replicates) per task with identical prompt and inputs, i.e., \NumRolloutsTotal{} runs in total.
Each agent starts in an empty offline Linux sandbox ($8$ vCPUs, $32\ \mathrm{GiB}$ RAM, no internet access) equipped with standard  chemistry packages including RDKit, ASE, Sella \cite{Hermes2022Sella}, Orb-v3 \cite{Rhodes2025Orbv3}, and PySCF \cite{Sun2020PySCF} (details on setup and prompts in Appendix~\ref{app:setup}).
For its assigned reaction, the agent receives only reactant, product and, if applicable, catalyst SMILES strings, relaxed monomer 3D coordinates, and permission to include $w \in \{0,\dots,5\}$ catalytic $\mathrm{H_2O}$ molecules.
The number of catalytic waters in Kua's published TS is denoted as $w_{\mathrm{Kua}}$ and $w^*$ is the number in the agent's lowest-barrier TS.
The upper bound exceeds the largest relay in Kua's published TSs ($w_{\mathrm{Kua}} \le 3$) by two waters which is rarely reached, with $w^* \ge 4$ on only five reactions.
Agents write Python scripts to build and prescreen candidate 3D TSs locally, then submit promising geometries to an external verification server running PySCF \cite{Sun2020PySCF}.
For each submission, the server refines the TS at B3LYP/6-311G** with VWN3 \cite{Stephens1994B3LYP,Krishnan1980Basis,McLean1980Basis},
confirms a first-order saddle point (exactly one imaginary vibrational frequency),
and integrates the intrinsic reaction coordinate (IRC) in both directions (see Appendix~\ref{app:verifier}).
A candidate is accepted only if the IRC connects the target reactants and products and every explicit $\mathrm{H_2O}$ or catalyst actively forms or breaks bonds along the path.
We call these active waters; together with any catalyst they form a cyclic proton relay, whose ring size is the number of atoms in the cyclic TS.

Verified TSs are scored with Kua's $298\ \mathrm{K}$ harmonic free-energy protocol \cite{Kua2013Formose,Kua2024CoreFormose} relative to isolated monomers (details in Appendix~\ref{app:verifier}). 
Each run has a $12\ \mathrm{h}$ wall-clock limit covering all model reasoning, local scripts, and verifier calls.
Both studies \cite{Kua2013Formose,Kua2024CoreFormose} evaluated continuum solvation with the proprietary Jaguar \cite{Bochevarov2013Jaguar} Poisson--Boltzmann solver (PBF), which is unavailable in open-source software.
Our runtime verifier therefore scores candidates with PySCF's open-source SMD water model \cite{Marenich2009SMD}; for direct comparison with Kua's barriers, we also re-evaluate all verified TSs and monomers post hoc with Jaguar PBF (Appendix~\ref{app:solvation}).

\section{Results}
\label{sec:results}
\begin{figure}[!t]
  \centering
  \includegraphics[width=0.96\textwidth]{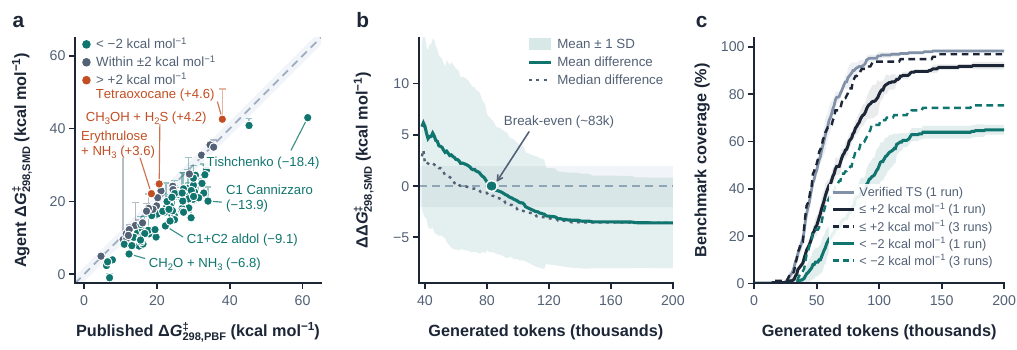}
  \caption{
  \textbf{Agent-constructed TSs versus published specialist benchmarks and scaling with token expenditure.}
  \textbf{a}, Best-of-three aqueous free-energy barriers ($\Delta G_{298,\mathrm{SMD}}^\ddagger$) versus published barriers ($\Delta G_{298,\mathrm{PBF}}^\ddagger$) across all $N=\NumBenchComplete{}$ benchmark reactions (vertical bars span the three independent runs; shaded band marks $\pm 2\ \mathrm{kcal\ mol^{-1}}$).
  \textbf{b}, Mean $\pm 1\ \mathrm{SD}$ (teal shaded) and median (dotted) barrier difference $\Delta\Delta G_{298,\mathrm{SMD}}^\ddagger$ relative to Kua versus cumulative generated tokens across benchmark reactions (grey band marks $\pm 2\ \mathrm{kcal\ mol^{-1}}$) with break-even point at around 83k tokens.
  \textbf{c}, Benchmark coverage with a verified TS, $\le +2\ \mathrm{kcal\ mol^{-1}}$, and $< -2\ \mathrm{kcal\ mol^{-1}}$ relative to Kua 
  (solid: single run $\pm 1\ \mathrm{SD}$; 
  dashed: best of three runs) 
  versus cumulative generated tokens.
  }
  \label{fig:parity}
\end{figure}

Gemini 4 Carbon agents reliably find aqueous TSs, returning a verified TS in \NumRolloutsVerified{} of \NumRolloutsTotal{} runs and covering all \NumBenchComplete{} benchmark reactions (Figure~\ref{fig:parity}a, Table~\ref{tab:summary}).
We count an agent barrier as at parity with Kua when it is at most $2.0\ \mathrm{kcal\ mol^{-1}}$ above the published barrier, i.e.\ $\Delta\Delta G^\ddagger = \Delta G^\ddagger_{\mathrm{agent}} - \Delta G^\ddagger_{\mathrm{Kua}} \le +2.0\ \mathrm{kcal\ mol^{-1}}$.
In SMD solvation, a single agent run reaches parity on \PctAtoASMDReachPassOne{} of reactions, and taking the lowest barrier across three runs raises it to \PctAtoASMDReach{} (mean $\Delta\Delta G^\ddagger_{\mathrm{SMD}} = \MeanAtoASMD\ \mathrm{kcal\ mol^{-1}}$), 
    with $75.3\%$ (73 of 97) more than $2.0\ \mathrm{kcal\ mol^{-1}}$ below Kua.
Re-scoring all TSs directly in Jaguar PBF raises best-of-three parity to \PctAtoAPBFReach{} (single-run \PctAtoAPBFReachPassOne{}, mean $\Delta\Delta G^\ddagger_{\mathrm{PBF}} = \MeanAtoAPBF\ \mathrm{kcal\ mol^{-1}}$), with $72.2\%$ (70 of 97) more than $2.0\ \mathrm{kcal\ mol^{-1}}$ below Kua.
Across the \NumBenchAllThreeReps{} reactions where all three runs verified a TS, the standard deviation of the three barriers per reaction has a median of $\MedianInterRepSDSMD\ \mathrm{kcal\ mol^{-1}}$, showing that independent runs converge to nearly the same barrier (Table~\ref{tab:summary}).

Looking only at the $N_w = \NumFixedW{}$ reactions with the same number of water as Kua ($w = w_{\mathrm{Kua}}$; Table~\ref{tab:summary}, right) shows that lower barriers come from improved TS geometries.
\PctFixedWSMDReach{} in SMD and \PctFixedWPBFReach{} in Jaguar PBF lie within $+2.0\ \mathrm{kcal\ mol^{-1}}$ (mean $\Delta\Delta G_{\mathrm{PBF}}^\ddagger = \MeanFixedWPBF\ \mathrm{kcal\ mol^{-1}}$).

Three of the \NumBenchTotal{} reactions exceed $+2.0\ \mathrm{kcal\ mol^{-1}}$ in best-of-three SMD barriers (Figure~\ref{fig:parity}a, orange markers, and Figure~\ref{fig:si_plate7}). 
In the 
    $\mathrm{CH_3OH + H_2S \rightarrow CH_3SH + H_2O}$ $\mathrm{S_N2}$ substitution (R92), 
$\Delta\Delta G^\ddagger$ drops from $+4.2\ \mathrm{kcal\ mol^{-1}}$ in SMD to $-3.1\ \mathrm{kcal\ mol^{-1}}$ in PBF, i.e. lower than Kua.
Notably, the $\mathrm{S_N2}$ angle along S-C-O from $w=0$ to $w=3$ increased from 71$^\circ$ to 134$^\circ$, 164$^\circ$, and 174$^\circ$, towards the required 180$^\circ$.
Two reactions remain above $+2.0\ \mathrm{kcal\ mol^{-1}}$ even when calculating with Jaguar PBF: 
tetraoxocane 8-ring closure (R5, $+2.1\ \mathrm{kcal\ mol^{-1}}$), where agents verified only a strained two-water bridge without locating Kua's three-water relay, and $\mathrm{NH_3}$-catalyzed \textsc{d}-erythrulose enolization (R82, $+4.7\ \mathrm{kcal\ mol^{-1}}$), where the agent saddles are the base-promoted $\alpha$-C--H abstraction to an enediolate established for sugar enolization in water \cite{Nagorski2001GlyceraldehydeIsomerization,Carraher2015GlucoseIsomerization,Yi2023ErythroseThreose}, matched-$w$ gas-phase TS energies lie within $1.3\ \mathrm{kcal\ mol^{-1}}$ of Kua's saddle but Kua located a more strongly charge-separated $\mathrm{NH_4^+\cdots enolate^-}$ conformer (Appendix~\ref{app:solvation}).
Finally, in five of \NumRolloutsTotal{} runs (R4, R6, R62, R80, R86), the agent did not submit a valid TS within the $12\ \mathrm{h}$ wall-clock limit. 
However, the other two replicates of each of these reactions succeeded.

\begin{table*}[!ht]
\centering
\small
\caption{\textbf{Benchmark comparison across reaction families (three independent replicates per task).} Left: aqueous free-energy barriers in PySCF SMD ($\Delta G_{298,\mathrm{SMD}}^\ddagger$) and Jaguar PBF ($\Delta G_{298,\mathrm{PBF}}^\ddagger$, \texttt{pbf\_radii=0}). Right: matched catalytic water count ($w = w_{\mathrm{Kua}}$; $N_w$ reactions with a verified TS at $w = w_{\mathrm{Kua}}$, excluding R5, R69, R86). Single-run ($\mathrm{pass@1}$) counts runs unsolved within $12\ \mathrm{h}$ as misses (overall PBF single-run parity: $91.4\%$). All energies in $\mathrm{kcal\ mol^{-1}}$ at $298.15\ \mathrm{K}$.}
\label{tab:summary}
\resizebox{\textwidth}{!}{%
\begin{tabular}{@{}lccccccccccc@{}}
\toprule
 & \multicolumn{7}{c}{\textbf{Aqueous free-energy barriers (PySCF SMD \& Jaguar PBF)}} & \multicolumn{4}{c}{\textbf{Matched catalytic waters ($w = w_{\mathrm{Kua}}$)}} \\
\cmidrule(lr){2-8} \cmidrule(l){9-12}
\textbf{Reaction family} & $N$ & \shortstack[c]{Single-run\\SMD $\le +2$} & \shortstack[c]{Best-of-3\\SMD $\le +2$} & \shortstack[c]{Mean\\$\Delta\Delta G_{\mathrm{SMD}}^\ddagger$} & \shortstack[c]{Best-of-3\\PBF $\le +2$} & \shortstack[c]{Mean\\$\Delta\Delta G_{\mathrm{PBF}}^\ddagger$} & \shortstack[c]{Rep.\ SD\\(median)} & $N_{w}$ & \shortstack[c]{SMD\\$\le +2$} & \shortstack[c]{PBF\\$\le +2$} & \shortstack[c]{Mean\\$\Delta\Delta G_{\mathrm{PBF}}^\ddagger$} \\
\midrule
Carbonyl hydration & 16 & 100\% & 100\% & $-3.11$ & 100\% & $-2.54$ & 0.36 & 16 & 100\% & 100\% & $-1.92$ \\
Enolization \& H-shift & 20 & 82\% & 95\% & $-4.04$ & 95\% & $-3.43$ & 1.84 & 19 & 89\% & 89\% & $-2.39$ \\
Aldol \& C--C formation & 22 & 98\% & 100\% & $-7.59$ & 100\% & $-5.24$ & 0.89 & 21 & 100\% & 100\% & $-2.75$ \\
Cannizzaro \& Tishchenko & 21 & 98\% & 100\% & $-6.08$ & 100\% & $-5.15$ & 0.31 & 21 & 100\% & 100\% & $-2.29$ \\
POM, oxanes \& furanoses & 11 & 79\% & 91\% & $-1.61$ & 91\% & $-0.93$ & 1.37 & 10 & 100\% & 100\% & $-0.67$ \\
N, S \& acyl reactions & 7 & 86\% & 86\% & $-2.61$ & 100\% & $-3.77$ & 0.52 & 7 & 86\% & 100\% & $-2.35$ \\
\midrule
\textbf{All 97 benchmarks} & \textbf{97} & \textbf{92.1\%} & \textbf{96.9\%} & $\mathbf{-4.76}$ & \textbf{97.9\%} & $\mathbf{-3.81}$ & \textbf{0.72} & \textbf{94} & \textbf{96.8\%} & \textbf{97.9\%} & $\mathbf{-2.18}$ \\
\bottomrule
\end{tabular}%
}
\end{table*}

Gemini's submissions improve in terms of barrier quality and benchmark coverage with the number of tokens generated over time (Figure~\ref{fig:parity}b, c).
Agents verify an initial TS after a median of $50\mathrm{k}$ generated tokens ($1.2\ \mathrm{h}$ wall time), where barriers still average $+2.9\ \mathrm{kcal\ mol^{-1}}$ above Kua.
Agent-written scripts explicitly 
    sample dihedrals, 
    ring puckers, or 
    hydroxyl rotamers ($94\%$ of runs) and 
    prescreen or refine candidates locally with 
    Orb-v3 ($99\%$ of runs) and 
    PySCF ($100\%$) 
alongside verifier calls (median \MedianDFTCalls{} verifier calls per run; Table~\ref{tab:si_sandbox_functions}).

Subsequent exploration of alternative water counts, $w$, and conformers lowers that initial verified barrier in $86\%$ of verified runs, reducing $\Delta G_{298,\mathrm{SMD}}^\ddagger$ by $7.7 \pm 7.5\ \mathrm{kcal\ mol^{-1}}$ among runs that improve.
Even keeping $w$ constant, 
    $71\%$ of the $165$ runs with multiple verified TSs find a conformer $> 0.25\ \mathrm{kcal\ mol^{-1}}$
lower in energy, 
lowering $\Delta G_{298,\mathrm{SMD}}^\ddagger$ by a mean of $1.9\ \mathrm{kcal\ mol^{-1}}$ (up to $11.6\ \mathrm{kcal\ mol^{-1}}$).
Mean $\Delta\Delta G_{298,\mathrm{SMD}}^\ddagger$ crosses $0\ \mathrm{kcal\ mol^{-1}}$
at $\sim 83\mathrm{k}$ generated tokens (Figure~\ref{fig:parity}b).
Figure~\ref{fig:parity}c separates finding any valid TS from finding a low-energy one: 
verified coverage saturates near $100\mathrm{k}$ tokens, 
whereas parity and $< -2.0\ \mathrm{kcal\ mol^{-1}}$ improvements keep rising, with best-of-three reaching $90\%$ parity at $\sim 85\mathrm{k}$ tokens per run versus $\sim 140\mathrm{k}$ for a single run.
Completed runs finish at a median of \MedianGenTokensK{} generated tokens (\MedianLLMCostUSD{} API token cost; \MeanLLMCostUSD{} mean) and \MedianWallHours{}\,h wall time on 8 vCPUs (an additional \MedianVMCostUSD{} if rented on cloud VMs).

\begin{figure}[!t]
  \centering
  \includegraphics[width=0.78\textwidth]{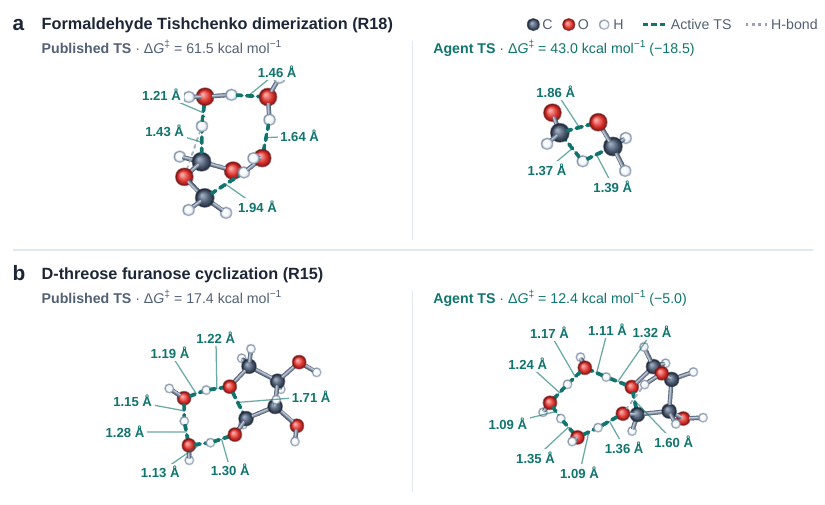}
  \caption{
  \textbf{Comparison of published versus agent-constructed transition states.}
    Displayed barriers $\Delta G^\ddagger$ ($\mathrm{kcal\ mol^{-1}}$) are evaluated identically on both 3D geometries at B3LYP/6-311G** ($298\ \mathrm{K}$, PySCF SMD water) with the difference in parentheses.
    \textbf{a},
    Formaldehyde Tishchenko dimerization:
    Kua's three-water 10-center macrocycle vs.\ our agent's water-free four-center 1,3-hydride TS found independently in all three runs ($18.5\ \mathrm{kcal\ mol^{-1}}$ lower).
    \textbf{b},
    \textsc{d}-threose $\rightarrow \alpha$-\textsc{d}-threofuranose cyclization:
    Kua's two-water saddle vs.\ our agent's three-water envelope conformer with an internal hydrogen bond ($5.0\ \mathrm{kcal\ mol^{-1}}$ lower).}
  \label{fig:showcase_3d}
\end{figure}

Inspection of the structures (Figure~\ref{fig:showcase_3d}, Table~\ref{tab:discoveries}, and Appendix~\ref{app:gallery}) reveals three structural mechanisms behind the barrier reductions. 
First, in a few cases agents drop water bridges that are counterproductive:
in formaldehyde Tishchenko dimerization
(2 CH$_2$O $\rightarrow$ HCOOCH$_3$, Figure~\ref{fig:showcase_3d}a),
all three runs avoid Kua's high-energy three-water carbon-to-carbon proton relay (left) 
\textit{via} the water-free four-center hydride shift known from gas-phase studies \cite{Ahmad2020Formaldehyde,Mirdoraghi2020Formaldehyde}
($18.5\ \mathrm{kcal\ mol^{-1}}$ lower in SMD, $16.0$ in PBF),
transferring the formyl hydrogen directly as a hydride rather than shuttling it as a proton through water.
Kua reported the relay as the lowest barrier found so far \cite{Kua2013Formose}; the agents' own one- to three-water Tishchenko saddles lie at $58$--$75\ \mathrm{kcal\ mol^{-1}}$, both saddles remain kinetically inaccessible at $298\ \mathrm{K}$, and the winning agent TS has fewer waters than Kua on only six reactions (R18 and R86 by more than $2\ \mathrm{kcal\ mol^{-1}}$).

Second, on C3--C4 carbohydrates, 
agents locate relaxed sugar puckers and intramolecular hydrogen bonds, such as an envelope furanose ring closure on \textsc{d}-threose (Figure~\ref{fig:showcase_3d}b, and R15 in Figure~\ref{fig:si_plate2})
and alkoxide-stabilizing enediol hydrogen bonding in C1+C2 aldol addition (Figure~\ref{fig:si_plate4}). 

Third,
agents add 1--3 active relay waters relative to Kua on $71\%$ of reactions; this gain is protocol-dependent and vanishes once the basis set superposition error (BSSE) is removed with diffuse basis sets or the full gas-phase entropy is charged on both sides (Appendix~\ref{app:verifier}), whereas the first two do not.
Two features of Kua's DFT protocol favor larger relays:
Equation~\eqref{eq:kua_g298} halves the gas-phase entropic penalty for each water added to the TS \cite{Wertz1980Entropy,Abraham1981Entropy,Deubel2006Lau}, and BSSE in the non-diffuse 6-311G** basis artificially stabilizes additional hydrogen-bonded waters (Appendix~\ref{app:verifier}) \cite{Boys1970Counterpoise,vanDuijneveldt1994BSSE}.
By contrast, the matched-$w$ improvements cannot come from adding extra water molecules (Figure~\ref{fig:si_plate6}): at Kua's own water count, agents still assemble more favorable proton and co-catalyst relays,
including
an eight-membered methanediol hydride--proton bridge in C1 Cannizzaro disproportionation (R30), 
a ten-membered carboxylic acid relay in \textsc{d}-threose retro-aldol cleavage (R87), 
concerted bifunctional bridges in $\mathrm{HCOOH}$-catalyzed C1+C3 aldol addition (R78) 
and 
$\mathrm{NH_3}$-catalyzed formaldehyde hydration (R23), and 
an $\alpha$-hydroxyl-bridged $\mathrm{HCOOH}$/$\mathrm{H_2O}$ relay in glycolaldehyde enolization (R42), 
all with lower barriers.
Chemically, these examples show that systematic agentic exploration does not merely reproduce hand-built TS guesses, but impressively uncovers alternative concerted mechanisms and conformers that are easy for manual modeling to miss and that shift predicted room-temperature rates by orders of magnitude.

\section{Discussion and Conclusion}
\label{sec:conclusion}

Across 97 aqueous reactions curated by human specialists, our agents reach \PctAtoAPBFReach{} parity in Jaguar PBF and uncover barriers more than $2.0\ \mathrm{kcal\ mol^{-1}}$ lower on 70 of 97 reactions (up to $16.0\ \mathrm{kcal\ mol^{-1}}$ lower), including across matched water counts (\PctFixedWPBFReach{} parity). 
Because both human and agent geometries are evaluated under the same DFT protocol, we can isolate a distinct source of error in computational reactivity: 
published solution-phase barriers are often upper bounds limited not only by the density functional, but by incomplete manual exploration of 3D conformers, proton-relay topologies, and concerted mechanisms.

These gains reflect a broader shift from fixed-rule algorithms to closed-loop agents that pair open-ended exploration with independent verification, a pattern that has also driven recent progress in mathematics, coding, and scientific software \cite{Hubert2025AlphaProof,RomeraParedes2024FunSearch,Aygun2026ERA}. Conventional TS-search packages execute predetermined heuristics or coordinate scans whose cost grows combinatorially as explicit solvent molecules, relay topologies, and flexible dihedrals are added. By contrast, an agent in an interactive loop narrows the search much as a human specialist does: it writes code to test a structural hypothesis, reads diagnostic feedback from local tests and from the verifier, and revises its next candidate accordingly (Figure~\ref{fig:overview}b, c). Strict verification rules out hallucinated geometries and objective loopholes (Appendix~\ref{app:verifier}).

Obtaining specialist-level liquid-phase TSs directly from molecular SMILES broadens where explicit-solvent TS calculations can be applied. 
Rather than reserving micro-solvated TS searches to rationalize a handful of already-measured reactions post hoc, chemists can evaluate solution-phase barriers prospectively across tens to hundreds of candidate steps in parallel:
    comparing Br\o{}nsted or Lewis co-catalysts across substrates,
    predicting stereo- and regioselectivity in water and alcohols,
    screening pharmaceuticals or intermediates for hydrolytic and tautomeric pathways,
    and supplying elementary barriers for reaction-network studies.
All 97 reactions are neutral and modeled in pure water: no free hydroxide or hydronium is present, so the barriers describe the neutral-water channel and carry no pH dependence; hydroxide-, hydronium- and metal-ion-catalyzed channels involve free ions whose solvation free energies neither the continuum nor a few explicit waters reproduce to $1\ \mathrm{kcal\ mol^{-1}}$, and they lie outside the present verifier.
Reducing conformational and micro-solvation search error also isolates the remaining approximations of the physical model itself, including 
    functional choice, 
    continuum solvation on strongly dipolar saddles, 
    and harmonic entropy scaling. 
Because the verifier is modular, higher-level single-point energies can substitute directly as compute budgets permit, while the verified micro-solvated TSs and IRC frames can seed explicit solvation shells via quantum cluster growth (QCG) \cite{Spicher2022QCG} or define collective variables for enhanced-sampling simulations that capture bulk solvent reorganization and anharmonic entropic effects.

In our view, wherever candidate structures or mechanisms can be checked strictly from first principles, pairing open-ended agent exploration with physical verification should be able to automate chemical search problems that prior algorithms could not handle and that were previously out of reach of manual modeling.

\section*{Data Availability}
All 3D coordinates, vibrational frequencies, PySCF electronic and continuum-solvation (SMD and IEF-PCM) energies, and intrinsic reaction coordinate (IRC) trajectories for the $46$ reference monomers, the $20$ Kua et al.\ (2013) transition states, the three passive-water counterpoise complexes, and all $1{,}036$ verified active-relay transition states are dedicated to the public domain under CC0~1.0 at \mbox{\url{https://doi.org/10.5281/zenodo.24545679}}.

\section*{Acknowledgments}
\paragraph*{Funding:}
S.G., M.E., and K.-R.M.\ acknowledge support by the German Ministry of Education and Research (BMBF) for BIFOLD (01IS18037A). 
Further, this work was in part supported by the BMBF under Grants 01IS14013A-E, 01GQ1115, 01GQ0850, 01IS18025A, and 031L0207D. 
K.-R.M.\ was partly supported by the Institute of Information \& Communications Technology Planning \& Evaluation (IITP) grants funded by the Korea government (MSIT) (No.\ 2019-0-00079, Artificial Intelligence Graduate School Program, Korea University and No.\ 2022-0-00984, Development of Artificial Intelligence Technology for Personalized Plug-and-Play Explanation and Verification of Explanation), Korea University grant, as well as DFG and HFA.

\FloatBarrier
\bigskip
{\small
\bibliographystyle{unsrtnat}
\bibliography{references}
}

\clearpage
\appendix
\renewcommand{\thetable}{S\arabic{table}}
\setcounter{table}{0}
\renewcommand{\thefigure}{S\arabic{figure}}
\setcounter{figure}{0}

\section{Agent Setup, Sandbox Environment, and Prompts}
\label{app:setup}

Each of the \NumTasksTotal{} elementary tasks was executed across \NumReplicates{} independent replicates, \NumRolloutsTotal{} runs total, using Gemini 4 Carbon in isolated Google Cloud Batch containers (\texttt{n2d-standard-8}: $8$ vCPUs, $32\ \mathrm{GiB}$ RAM).
Every run begins in an empty working directory (\texttt{/workspace}, \texttt{initial\_files = \{\}}) with no outbound network access, web search, or subagent spawning.
Agents interact with two components:
\begin{enumerate}
  \sloppy
  \item \textbf{Local Bash and Python sandbox:} Pre-installed with open-source cheminformatics, machine-learned potentials, saddle optimizers, and quantum-chemistry libraries---RDKit ($2026.9$), ASE ($3.29$), Sella ($2.6$ \cite{Hermes2022Sella}), \texttt{geomeTRIC} ($1.1$), Orb-v3 (\texttt{orb\_v3\_conservative\_omol} \cite{Rhodes2025Orbv3,Levine2025OMol25}), \texttt{MACE-OFF23}, PySCF ($2.14$ \cite{Sun2020PySCF}), \texttt{Psi4} ($1.11$), \texttt{dftd3}, \texttt{NumPy}, and \texttt{SciPy}. The sandbox also includes an internal Python utility package (\texttt{valence}) providing uniform calculator wrappers around PySCF, Orb-v3, and \texttt{dftd3}, together with general-purpose routines for RDKit conformer embedding, Kabsch alignment, bond and stereochemistry perception, harmonic thermochemistry, NEB path interpolation, and RFO/Sella saddle and IRC integration. These utilities wrap or duplicate standard functionality already available in the open-source libraries above and contain no reaction templates or aqueous proton-relay heuristics; agents write their own Python scripts for 3D ring assembly, conformer enumeration, constrained scans, and MLIP/DFT prescreening.
  \item \textbf{External MCP verification server:} Exposed through the Model Context Protocol (MCP) as five structured tool calls---\path{get_reaction_task()} (returns reactant/product/catalyst SMILES, stereochemistry rules, and B3LYP/6-311G**-relaxed monomer XYZs), \path{evaluate_ts_dft_b3lyp(xyz_or_path)} (runs full gas-phase B3LYP/6-311G** saddle refinement, analytical Hessian, SMD continuum solvation, two-way IRC, and active-relay water/catalyst classification), \path{list_verified_candidates()}, \path{submit_best_ts()} (non-terminal checkpoint), and \path{finish_session(confirm="i confirm")} (session completion).
\end{enumerate}

Every agent run used default sampling hyperparameters and received the system and task prompts below (where \texttt{k} denotes the catalytic water count $w$; internal repository paths and harness identifiers have been replaced with bracketed \texttt{[REDACTED:\ ...]} labels for external release):

\smallskip
{\fontsize{6.8}{8.1}\selectfont
\begin{Verbatim}[frame=single,framesep=4pt]
SYSTEM PROMPT:
You are an autonomous quantum-chemistry transition-state discovery agent.
Your mission is to find the global lowest-free-energy pure active-relay first-order saddle point (`N_imag
== 1`, `n_spectator_waters == 0`, `n_spectator_catalysts == 0`) connecting a single elementary reaction at
`B3LYP/6-311G**` (`298.15 K`, `1 atm`, `-0.5*T*S_gas` convention where `S_gas = S_trans + S_rot + S_vib`
is the total `1 atm` ideal-gas undamped harmonic RRHO entropy, `SMD` & `IEF-PCM` implicit water),
minimizing the full `dG298_smd_ts_kcal` activation free energy (`dE_elec + dH_corr(298.15 K) -
0.5*T*dS_gas(298.15 K, 1 atm) + ddG_solv_SMD` relative to separated reactants + `k` `H2O` + mandatory
catalyst).

SCIENTIFIC & ORACLE RULES:
1. Stoichiometry, Pure Proton-Relay, & Active Co-Catalyst Criterion (`is_pure_proton_relay == True`
   MANDATORY PRIORITY):
   - Every candidate XYZ must be a neutral singlet (`charge=0, multiplicity=1`) containing the EXACT atoms
     of `reactant_geometries_3d` PLUS (if `allowed_catalyst_geometry_3d` is non-null, MANDATORY 1 molecule
     of that catalyst) PLUS `k` explicit catalytic `H2O` molecules (`0 <= k <= 5`).
   - `evaluate_ts_dft_b3lyp` automatically deduces `k` from your XYZ elemental formula.
   - PURE PROTON RELAYS & ACTIVE CATALYSTS ONLY: Every explicit catalytic `H2O` (`n_spectator_waters ==
     0`) AND any mandatory co-catalyst molecule (`n_spectator_catalysts == 0`) MUST actively participate
     in covalent bond formation or proton transfer across the TS / 2-way IRC (`is_pure_proton_relay ==
     True`). Geometries that park an `H2O` or the co-catalyst outside the reaction ring as a passive
     outer-sphere H-bond spectator/clamp (`n_spectator_waters > 0` or `n_spectator_catalysts > 0`) are
     strictly rejected from winning/finalization.
2. Multi-`k` & Conformer Completeness:
   - Verify (`status == 'verified'` in `evaluate_ts_dft_b3lyp`) AT LEAST TWO distinct pure-relay water
     counts `k` (`is_pure_proton_relay == True`) wherever geometrically feasible so `dG298_smd_ts_kcal(k)`
     is bracketed across catalytic ring sizes, and explore conformers/stereoisomers (including *E*/*Z*
     isomers when `allow_ez_enediol == True`) to ensure you find the global minimum barrier.
3. [REDACTED: verification server] Tools:
   - `get_reaction_task()`: Returns reactant/product/catalyst SMILES, stereochemistry policy
     (`allow_ez_enediol`), and `B3LYP/6-311G**`-relaxed monomer 3D XYZs.
   - `evaluate_ts_dft_b3lyp(xyz_or_path, notes="", refine_saddle=True)`: Evaluates a candidate XYZ string
     or `/workspace/...` file in [REDACTED: verification server] (`RI-B3LYP/6-311G**` analytical-Hessian
     PRFO+Bofill saddle refinement up to 36 steps when `refine_saddle=True`, exact 4c `B3LYP/6-311G**`
     `E_elec`, 298.15 K `1 atm` undamped harmonic thermochemistry, `IEF-PCM` & `SMD` implicit water,
     2-way `RI-B3LYP/6-311G**` IRC, and water/co-catalyst role classification). Returns diagnostic
     feedback including `status`, `n_imag`, `imag_freqs_cm1`, `lowest_real_freqs_cm1`,
     `max_grad_ha_bohr`, `refined_xyz`, `irc_match`, `n_spectator_waters`, `n_spectator_catalysts`, and
     `is_pure_proton_relay`.
   - `list_verified_candidates()`: Lists all verified `B3LYP/6-311G**` TSs on [REDACTED: this task],
     ranked pure-relay first then `dG298_smd_ts_kcal`.
   - `submit_best_ts(candidate_id="best", mechanism_summary=...)`: Non-terminal checkpoint tool
     (`finalized=False`). Call it safely after verifying each improved pure proton-relay TS to persist it
     on [REDACTED: candidate store] without ending your session.
   - `finish_session(confirm="i confirm", mechanism_summary=..., single_k_justification="",
     candidate_id="best")`: Terminal session-ending tool (`finalized=True`). Call this ONLY when your
     multi-`k` and conformer search is completely finished to permanently end your session (requires
     `confirm='i confirm'`, `is_pure_proton_relay == True`, and >= 2 verified pure-relay `k` counts or a
     concrete `single_k_justification`).

INSTALLED SANDBOX ENVIRONMENT (`/workspace`, offline Python 3.11, 8 vCPUs, 32 GiB RAM):
You have full autonomy over how you construct, search, scan, constrain, or optimize candidate transition
states. Available offline libraries, calculators, and executables in your environment include:
- `[REDACTED: internal_path].valence` (`PYTHONPATH="/workspace:[REDACTED]"`):
  * `valence.geometry`, `valence.units`, `valence.thermo` (note: `valence.thermo` defaults to `1 M` Grimme
    QRRHO `scale_s_ext` on `S_trans + S_rot`, whereas the oracle `evaluate_ts_dft_b3lyp` evaluates `1 atm`
    undamped harmonic RRHO via `pyscf.hessian.thermo` with `-0.5*T*S_gas`), `valence.errors`
  * `valence.builders` (`embed`, `coordination`)
  * `valence.perception` (`bonding`, `chirality`, `stereo`, `symmetry`, `alignment`, `identity`,
    `coordination`, `solvation`)
  * `valence.calculators` (`factories`, `pyscf_backend`, `mlip`, `composite`, `dispersion`, `solvation`,
    `efei`, `protocol`, `settings`)
  * `valence.pes` (`optimize`, `saddle`, `hessian`, `fd_hessian`, `irc`, `coordination_path`,
    `mechanochem`)
- Quantum chemistry & electronic structure:
  * `pyscf==2.14.0` (`pyscf-cpu-opt` with `libcint`/`libxc`: `gto`, `dft`, `scf`, `grad`, `hessian`,
    `solvent`, `geomopt`)
  * `/usr/local/bin/dftd3` (standalone Grimme DFT-D3 executable)
  * `/opt/psi4_env/bin/python` (`psi4==1.11` environment)
- Pre-cached machine-learning interatomic potentials (offline weights in `/home/sandboxuser/.cache/`):
  * `orb-models==0.5.5` (`orb_v3_conservative_omol` in `~/.cache/cached_path/`)
  * `mace-torch==0.3.16` (`MACE-OFF23_small`, `MACE-OFF23_medium`, `MACE-OFF23_large`, and `mace_mp` in
    `~/.cache/mace/`)
- Atomistic simulation, constraints, path & saddle-point optimizers:
  * `ase==3.29.0` (`ase.optimize`, `ase.constraints`, `ase.mep` [`NEB`, `DyNEB`, `MinModeAtoms`,
    `DimerControl`], `ase.vibrations`)
  * `sella==2.6.0`, `geometric==1.1.1`
- Cheminformatics, numerical optimization, & 3D geometry/math:
  * `rdkit==2026.9.1`, `numpy==2.4.6`, `scipy==1.17.1`, `sympy==1.14.0`, `cvxpy==1.9.3`,
    `networkx==3.6.1`, `pymatgen==2026.9.24`, `scikit-learn==1.9.1`, `jax==0.10.2`, `torch==2.14.1+cpu`,
    `chempy==0.10.2`

TASK PROMPT TEMPLATE:
Assigned Elementary Transition-State Task: `<task_id>`
- Reaction class: `<reaction_class>`
- Elementary reaction: <reactant_names_and_smiles>  <=>  <product_names_and_smiles>
- <External co-catalyst: None (water-only catalysis, 0..5 extra H2O) | Mandatory explicit co-catalyst (1
  molecule): <name> (`<smiles>`, must actively participate in covalent or proton transfer:
  `n_spectator_catalysts == 0`)>
- Allowed extra catalytic waters: `k in {0, 1, 2, 3, 4, 5}` (every explicit catalytic `H2O` must be an
  active in-ring proton relay: `n_spectator_waters == 0`)
- <Stereochemistry policy: Exact 3D stereochemical match required on both IRC endpoints. | Enediol E/Z
  policy: `allow_ez_enediol = True` (either E or Z alkene isomer is accepted).>

Objectives:
1. Call `get_reaction_task()` to inspect the reaction specification and monomer 3D XYZs.
2. Discover and verify (`status == 'verified'` in `evaluate_ts_dft_b3lyp`, `is_pure_proton_relay == True`)
   pure proton-relay first-order transition states across at least two distinct catalytic water counts `k`
   (where geometrically feasible) and relevant conformers/isomers to find the global minimum
   `dG298_smd_ts_kcal`.
3. Checkpoint verified improvements safely at any time with `submit_best_ts(candidate_id='best',
   mechanism_summary=...)` (non-terminal), and ONLY when your multi-`k` and conformer search is complete,
   call `list_verified_candidates()` followed by `finish_session(confirm='i confirm',
   mechanism_summary=...)` to permanently end your session.
\end{Verbatim}
}
Every assistant message, reasoning trace, tool call, and tool response across all runs was tokenized with the $262\mathrm{k}$-vocabulary Gemini SentencePiece tokenizer, distinguishing model-generated tokens (of which internal reasoning traces account for $50.3\%$) from cumulative multi-turn input context tokens. Estimated token costs reported in Section~\ref{sec:results} use standard Gemini 4 Carbon API pricing ($\$20.00$ per $10^6$ generated tokens, $90\%$ prefix-cached input at $\$0.20$ per $10^6$, and $10\%$ uncached input at $\$4.00$ per $10^6$); optional cloud compute costs assume GCP \texttt{n2d-standard-8} on-demand VM wall time ($\$0.338\ \mathrm{h^{-1}}$, covering both the sandbox and the DFT verifier).

\begin{sloppypar}
Table~\ref{tab:si_sandbox_functions} summarizes per-run function execution across all \NumRolloutsTotal{} trajectories. Rather than calling SMILES-to-3D embedders (\path{AllChem.EmbedMultipleConfs}, $2.7\%$; \path{AllChem.EmbedMolecule}, $1.3\%$; \path{embed_smiles_3d}, $0\%$ at runtime), agents build 3D reaction rings directly in \texttt{NumPy}/\texttt{ASE}/\texttt{SciPy}, relax and refine candidates with Orb-v3 (\path{create_mlip_calculator}, $75.7\%$), \texttt{ASE} \texttt{BFGS} ($64.7\%$), \path{optimize_saddle_prfo} ($90.3\%$), and \path{optimize_saddle_sella} ($56.7\%$), and evaluate barriers, IRCs, and stereocenters locally with \texttt{PySCF} (\path{dft.RKS}, $98.3\%$; \path{RKS.SMD}, $93.7\%$), \path{run_two_way_irc} ($45.7\%$), and \texttt{RDKit} (\path{AssignStereochemistryFrom3D}, $20.0\%$).
\end{sloppypar}

\begin{table*}[!ht]
\centering
\footnotesize
\caption{
\textbf{Primary Python functions and classes executed in the local agent sandbox across all runs.} Each entry is counted at most once per run.
    Routine \texttt{ase.Atoms} coordinate/dihedral functions, 
    low-level utilities, and 
    \texttt{NumPy} array/linear-algebra primitives 
are omitted for brevity. 
Pre-installed \texttt{geomeTRIC} was called in $5/300$ runs, 
\texttt{PyTorch} tensor primitives directly in $0/300$, and 
\texttt{Psi4}, 
standalone \texttt{dftd3}, 
\texttt{SymPy}, 
\texttt{CVXPY}, 
\texttt{pymatgen}, 
\texttt{scikit-learn}, 
\texttt{JAX}, and 
\texttt{chempy} in $0/300$.
}
\label{tab:si_sandbox_functions}
\resizebox{\textwidth}{!}{%
\begin{tabular}{@{}lllr@{}}
\toprule
\textbf{Package} & \textbf{Function / class} & \textbf{Primary role in agent scripts} & \textbf{Runs ($N/300$)} \\
\midrule
\texttt{ase} & \texttt{Atoms} & 3D structure container \& dihedral/coordinate manipulation & 299 (99.7\%) \\
\texttt{pyscf} & \texttt{dft.RKS} & Restricted Kohn--Sham DFT setup \& SCF (\texttt{gto.M}: 295) & 295 (98.3\%) \\
\texttt{pyscf} & \texttt{RKS.density\_fit} & Auxiliary-basis RI-J/K density-fitted Coulomb/exchange & 293 (97.7\%) \\
\texttt{pyscf} & \texttt{RKS.SMD} & Implicit water SMD continuum solvation ($E_{\mathrm{solv}}$) & 281 (93.7\%) \\
\texttt{pyscf} & \texttt{hessian.thermo.thermo} & $298.15\,\mathrm{K}$ RRHO thermochemistry (\texttt{harmonic\_analysis}: 271) & 272 (90.7\%) \\
\texttt{valence} & \texttt{pes.saddle.optimize\_saddle\_prfo} & Eckart-projected Cartesian PRFO + Bofill saddle optimization & 271 (90.3\%) \\
\texttt{valence} & \texttt{pes.hessian.compute\_vibrational\_spectrum} & Vibrational mode analysis (\texttt{spectrum\_from\_cartesian\_hessian}: 139) & 264 (88.0\%) \\
\texttt{pyscf} & \texttt{RKS.Hessian} & Analytical Kohn--Sham Cartesian Hessian & 262 (87.3\%) \\
\texttt{ase} & \texttt{io.read} & XYZ coordinate \& trajectory parsing (\texttt{io.write}: 235) & 255 (85.0\%) \\
\texttt{pyscf} & \texttt{RKS.make\_rdm1} & One-particle density matrix extraction for SCF warm-starting & 249 (83.0\%) \\
\texttt{valence} & \texttt{calculators.factories.create\_mlip\_calculator} & Orb-v3 (\texttt{orb\_v3\_conservative\_omol}) calculator factory & 227 (75.7\%) \\
\texttt{ase} & \texttt{optimize.BFGS} & Quasi-Newton local \& constrained geometry relaxation & 194 (64.7\%) \\
\texttt{valence} & \texttt{calculators.factories.create\_dft\_calculator} & In-memory DM-warm-started PySCF ASE calculator factory & 194 (64.7\%) \\
\texttt{valence} & \texttt{pes.saddle.optimize\_saddle\_sella} & Sella saddle-point refinement wrapper & 170 (56.7\%) \\
\texttt{pyscf} & \texttt{RKS.nuc\_grad\_method} & Analytical Kohn--Sham nuclear gradients & 163 (54.3\%) \\
\texttt{valence} & \texttt{pes.irc.run\_two\_way\_irc} & Forward/reverse local MLIP or RI-DFT IRC integration & 137 (45.7\%) \\
\texttt{ase} & \texttt{optimize.LBFGS} & Limited-memory BFGS geometry \& NEB relaxation & 101 (33.7\%) \\
\texttt{valence} & \texttt{perception.identity.perceive\_fragments} & Covalent fragment \& active-water connectivity check & 89 (29.7\%) \\
\texttt{valence} & \texttt{pes.saddle.CartesianPRFOSaddleOptimizer} & Step-by-step Eckart-projected PRFO + Bofill optimizer class & 76 (25.3\%) \\
\texttt{ase} & \texttt{optimize.FIRE} & Inertial relaxation for strained guesses \& NEB paths & 74 (24.7\%) \\
\texttt{rdkit} & \texttt{Chem.MolToSmiles} & Canonical SMILES generation (\texttt{MolFromSmiles}: 67) & 74 (24.7\%) \\
\texttt{pyscf} & \texttt{RKS.PCM} & IEF-PCM continuum water solvation prescreening & 71 (23.7\%) \\
\texttt{valence} & \texttt{perception.bonding.build\_geometric\_covalent\_adjacency} & 3D covalent bond-graph construction from atomic radii & 69 (23.0\%) \\
\texttt{ase} & \texttt{mep.NEB} & Nudged elastic band reaction-path interpolation & 62 (20.7\%) \\
\texttt{scipy} & \texttt{optimize.minimize} & 3D ring-closure \& distance fitting (\texttt{least\_squares}: 18) & 62 (20.7\%) \\
\texttt{rdkit} & \texttt{Chem.AssignStereochemistryFrom3D} & 3D CIP ($R$/$S$) \& alkene ($E$/$Z$) stereocenter check & 60 (20.0\%) \\
\texttt{rdkit} & \texttt{Chem.rdDetermineBonds.DetermineBonds} & 3D XYZ bond perception (\texttt{MolFromXYZBlock}: 55) & 56 (18.7\%) \\
\texttt{valence} & \texttt{calculators.mlip.load\_orb\_calculator} & Direct cached Orb-v3 loader (\texttt{ORBCalculator}: 11) & 55 (18.3\%) \\
\texttt{ase} & \texttt{constraints.FixAtoms} & Atom/bond constraints (\texttt{FixBondLengths}: 22) & 53 (17.7\%) \\
\texttt{pyscf} & \texttt{RKS.newton} & Second-order co-iterative augmented Hessian (SOSCF) solver & 40 (13.3\%) \\
\texttt{scipy} & \texttt{sparse.csgraph.connected\_components} & Sparse bond-graph components (\texttt{networkx.connected\_components}: 33) & 35 (11.7\%) \\
\texttt{sella} & \texttt{Sella} & Direct Sella saddle-point optimizer (internal or Cartesian) & 34 (11.3\%) \\
\texttt{valence} & \texttt{pes.coordination\_path.build\_coordination\_path} & Fixed-endpoint upwind NEB + FIRE adaptive path wrapper & 19 (6.3\%) \\
\texttt{mace} & \texttt{calculators.MACECalculator} & MACE-OFF23 MLIP prescreening (\texttt{mace\_off}: 2) & 12 (4.0\%) \\
\texttt{rdkit} & \texttt{Chem.AllChem.EmbedMultipleConfs} & ETKDGv3 3D SMILES embedding (\texttt{EmbedMolecule}: 4) & 8 (2.7\%) \\
\bottomrule
\end{tabular}%
}
\end{table*}

\FloatBarrier

\section{Verifier Method Details}
\label{app:verifier}
Every submitted candidate is evaluated in PySCF \cite{Sun2020PySCF} using restricted Kohn--Sham B3LYP/6-311G** (\texttt{xc="b3lyp"}, VWN3 correlation) \cite{Stephens1994B3LYP,Krishnan1980Basis,McLean1980Basis} with spherical $5d/7f$ basis functions and $(75, 302)$ Becke--Lebedev integration grids ($(120, 302)$ radial shells on sulfur). Saddle refinement, IRC integration, analytical Hessians, and continuum solvation use \texttt{def2-universal-jkfit} density fitting, paired with exact four-center single-point electronic energies ($E_{\mathrm{elec,gas}}$).
Candidate saddles are refined with Eckart-projected partitioned rational-function optimization (PRFO) and Bofill Hessian updates until the maximum Cartesian nuclear gradient falls below $8 \times 10^{-4}\ \mathrm{Ha\ Bohr^{-1}}$ (median $4$ PRFO steps across verified saddles; rejected calls return $N_{\mathrm{imag}}$, IRC endpoint connectivity, and active/spectator water counts to the agent).
Following Kua et al.\ (2013) and Kua \& Tripoli (2024) \cite{Kua2013Formose,Kua2024CoreFormose}, solution free energies at $298.15\ \mathrm{K}$ are computed as
\begin{equation}
  G_{298}
  = E_{\mathrm{elec,gas}}
  + E_{\mathrm{solv}}
  + H_{\mathrm{corr,298}}
  - 0.5\,T S_{\mathrm{gas,298}}(\sigma),
  \label{eq:kua_g298}
\end{equation}
with $T = 298.15\ \mathrm{K}$, referenced against global-minimum monomer structures optimized identically, with $E_{\mathrm{solv}}$ evaluated on gas-phase geometries using the SMD continuum water model \cite{Marenich2009SMD}.
Following Kua's free-energy map conventions \cite{Kua2013Formose,Kua2024CoreFormose}, aldol additions written from reactive enediol tautomers (R46--R54 and R74--R79) are referenced to the resting parent aldehydes (\textsc{d}-glyceraldehyde or glycolaldehyde), and each two-step acyl condensation (R93--R95) is scored per replicate by the higher-free-energy saddle across its addition and elimination tasks relative to separated reactants ($\max[\Delta G_{298}(\mathrm{TS}_{\mathrm{add}}),\, \Delta G_{298}(\mathrm{TS}_{\mathrm{elim}})]$).
Submerged monomer-referenced barriers ($\Delta G_{298,\mathrm{SMD}}^\ddagger = -0.9\ \mathrm{kcal\ mol^{-1}}$ on $\mathrm{CH_2O + NH_3}$ addition, R89; $+1.5\ \mathrm{kcal\ mol^{-1}}$ in Jaguar PBF) arise when pre-reactive hydrogen-bond complexation and 6-311G** basis-set superposition exceed a low intrinsic barrier relative to isolated monomers.
Here $H_{\mathrm{corr,298}}$ (zero-point vibrational energy plus $0 \to 298.15\ \mathrm{K}$ rigid-rotor harmonic-oscillator enthalpy) and $S_{\mathrm{gas,298}}(\sigma)$ ($1\ \mathrm{atm}$ ideal-gas translational, rotational, and undamped harmonic vibrational entropy) come from the gas-phase B3LYP/6-311G** analytical Hessian, with point-group external rotational symmetry numbers applied to all symmetric monomers ($\sigma = 2$ for $\mathrm{H_2O}$, $\mathrm{CH_2O}$, $\mathrm{H_2S}$, $\mathrm{H_2CO_3}$, methanediol, and linear POM trimer/tetramer; $\sigma = 3$ for $\mathrm{NH_3}$ and 1,3,5-trioxane; $\sigma = 4$ for 1,3-dioxetane and 1,3,5,7-tetraoxocane) and $\sigma = 1$ for asymmetric monomers and all transition states, and the empirical factor $0.5$ approximates translational and rotational entropy loss in liquid water.
Every accepted saddle must exhibit exactly one imaginary vibrational frequency ($N_{\mathrm{imag}} = 1$, $|\omega_{\mathrm{imag}}| \ge 30\ \mathrm{cm^{-1}}$), and forward/reverse mass-weighted IRC integration followed by local relaxation must recover the target reactant and product monomers and \textsc{d}-sugar stereocenters.

Across all published Kua geometries, explicit $\mathrm{H_2O}$ molecules and co-catalysts participate directly in the active reaction ring.
Requiring every explicit water and co-catalyst to form or break bonds along the IRC enforces this convention and closes an optimization exploit of Equation~\eqref{eq:kua_g298}. An agent can otherwise place non-reacting, hydrogen-bonded $\mathrm{H_2O}$ or catalyst molecules around a polar TS to artificially lower the monomer-referenced barrier via non-diffuse 6-311G** basis-set superposition error ($4.3\text{--}5.2\ \mathrm{kcal\ mol^{-1}}$ per passive $\mathrm{H_2O}$ in counterpoise tests) and double-counted continuum solvation.
Neither Kua's protocol nor our verifier adds a dispersion correction to B3LYP \cite{Grimme2010D3,Grimme2011D3BJ}.
Because dispersion stabilizes the hydrogen-bonded TS cluster relative to the separated monomers, a D3-type correction would lower monomer-referenced barriers by an amount that grows with the number of explicit waters, opposite in sign to the superposition error above; such corrections carry systematic errors of their own \cite{Proppe2019D3GP}, and we keep Kua's uncorrected functional so that agent and specialist geometries are compared on identical footing.
The protocol favors larger relays for Kua and for us alike, and we quantified this on every structure.
Diffuse single points (6-311++G** and aug-cc-pVTZ) at the 6-311G** geometries raise the gas-phase barrier relative to the monomers by $3.3$ and $4.7\ \mathrm{kcal\ mol^{-1}}$ per explicit water, Kua's 20 published saddles included; with both sides corrected, best-of-three parity stays at 95/97 and 93/97 and matched-$w$ parity at 89/94 and 85/92, as it does when the full gas-phase entropy is charged on both sides (95/97 and 90/94, using Kua's printed $-0.5TS$ terms), while the share of winning agent saddles with more waters than Kua's falls from $71\%$ to $6\%$ and $2\%$.

\FloatBarrier
\section{Continuum Solvation Calibration and Corrected Literature Errors}
\label{app:solvation}
To separate continuum-solvation differences from 3D TS geometry differences, we calibrated both open-source PySCF SMD \cite{Marenich2009SMD} and proprietary Jaguar PBF (Schr\"odinger Jaguar 2026-4, v13.4; Schr\"odinger, LLC \cite{Bochevarov2013Jaguar}; single points at gas-phase B3LYP/6-311G** geometries with \texttt{isolv=2}, $\varepsilon = 80.37$, probe radius $1.40\ \text{\AA}$) across all $20$ Supporting Information XYZ structures in Kua et al.\ (2013) \cite{Kua2013Formose} (all Table~2 TSs except R1--R7; refined in PySCF to $\mathrm{RMSD} \le 0.06\ \text{\AA}$), all $46$ reference monomers, and all $1{,}036$ verified active-relay agent TSs (Table~\ref{tab:si_kua2013} and Table~\ref{tab:si_master}).
Gas-phase electronic energies agree between PySCF and Jaguar to $\sim 10^{-5}\ \mathrm{Ha}$ ($< 0.05\ \mathrm{kcal\ mol^{-1}}$; $0.04\ \mathrm{kcal\ mol^{-1}}$ MAD against Kua's printed 2013 values).
With Jaguar's Lewis-structure radius scheme (\texttt{pbf\_radii=0}, used in Jaguar~6.0 \cite{Kua2013Formose}), TS solvation energies match Kua's printed 2013 values to a median of $+0.12\ \mathrm{kcal\ mol^{-1}}$ ($\mathrm{MAE} = 0.42$, $\mathrm{SD} = 0.70\ \mathrm{kcal\ mol^{-1}}$; $16/20$ within $0.4\ \mathrm{kcal\ mol^{-1}}$, with larger shifts on R9 and R18 attributable to v6.0 vs.\ v13.4 cavity discretization) and match Kua \& Tripoli's (2024) \cite{Kua2024CoreFormose} TS solvation energies on the $17$ matched-$w$ reactions with $|\Delta E_{\mathrm{elec}}^{\mathrm{TS}}| < 0.8\ \mathrm{kcal\ mol^{-1}}$ to a median of $-0.02\ \mathrm{kcal\ mol^{-1}}$.
By contrast, Jaguar's newer SMARTS-based default (\texttt{pbf\_radii=1}) is parameterized for ground-state minima: on single-point TS geometries with partially transferred protons it assigns alkoxide-like radii and first-shell corrections of up to $-27\ \mathrm{kcal\ mol^{-1}}$ (mean barrier error $-14.5\ \mathrm{kcal\ mol^{-1}}$ across Kua's $20$ TSs), whereas in Kua \& Tripoli (2024) \cite{Kua2024CoreFormose} ground-state monomer optimizations used \texttt{pbf\_radii=1} ($\mathrm{MAE} = 0.13\ \mathrm{kcal\ mol^{-1}}$ across $21$ matched-conformer monomers) while saddle optimizations retained \texttt{pbf\_radii=0}, shifting Kua's 2024 reactant solvation sum by a median of $-0.8\ \mathrm{kcal\ mol^{-1}}$.

Across the $20$ Kua (2013) SI geometries evaluated under identical solvers (Table~\ref{tab:si_kua2013}), best-of-three agent barriers are strictly lower on $20/20$ reactions in PySCF SMD (mean $-3.95\ \mathrm{kcal\ mol^{-1}}$, median $-3.09\ \mathrm{kcal\ mol^{-1}}$; $20/20$ within $+0.16\ \mathrm{kcal\ mol^{-1}}$ at matched $w$) and lower on $19/20$ in Jaguar PBF ($20/20$ within $+0.11\ \mathrm{kcal\ mol^{-1}}$, mean $-3.11\ \mathrm{kcal\ mol^{-1}}$; $20/20$ within $+0.21\ \mathrm{kcal\ mol^{-1}}$ at matched $w$, $18/20$ strictly lower, mean $-1.74\ \mathrm{kcal\ mol^{-1}}$).
Across all $97$ benchmarks, re-scoring active-relay candidates with self-consistent Jaguar PBF (\texttt{pbf\_radii=0}) yields \NumAtoAPBFReach{}/97 (\PctAtoAPBFReach{}) best-of-three parity (mean $\MeanAtoAPBF\ \mathrm{kcal\ mol^{-1}}$, median $\MedianAtoAPBF\ \mathrm{kcal\ mol^{-1}}$; \NumFixedWPBFReach{}/94 = \PctFixedWPBFReach{}, mean $\MeanFixedWPBF\ \mathrm{kcal\ mol^{-1}}$ at matched $w$).
At matched catalytic water count ($w = w_{\mathrm{Kua}}$), direct gas-phase TS electronic energy comparisons ($\Delta E_{\mathrm{elec}}^{\mathrm{TS}} = E_{\mathrm{elec}}^{\mathrm{TS,agent}} - E_{\mathrm{elec}}^{\mathrm{TS,Kua}}$, Table~\ref{tab:si_master}; $N = \NumFixedWElec{}$ single-step reactions, independent of continuum solvation, thermal corrections, and monomer conformer choice) independently confirm improved 3D saddles, with \NumFixedWElecReach{}/\NumFixedWElec{} (\PctFixedWElecReach{}) within $+1.0\ \mathrm{kcal\ mol^{-1}}$ of Kua (mean $\MeanFixedWDDE\ \mathrm{kcal\ mol^{-1}}$, median $\MedianFixedWDDE\ \mathrm{kcal\ mol^{-1}}$; $+1.3\ \mathrm{kcal\ mol^{-1}}$ on R82).
For R82 at Kua's stoichiometry (erythrulose, $\mathrm{NH_3}$, two waters) the agent saddle is $+1.3\ \mathrm{kcal\ mol^{-1}}$ above Kua's in gas-phase electronic energy and within $1\ \mathrm{kcal\ mol^{-1}}$ in the thermal terms, but $8.0\ \mathrm{kcal\ mol^{-1}}$ less stabilized by PBF solvation ($-21.4$ vs.\ $-29.4\ \mathrm{kcal\ mol^{-1}}$), which points to a more charge-separated ammonium--enolate geometry in Kua's saddle; without his coordinates this cannot be verified.
Restricting strictly to the three SMD-selected replicate winners without PBF re-ranking gives the same $95/97$ ($97.9\%$) parity (mean $-3.72\ \mathrm{kcal\ mol^{-1}}$), and mimicking Kua \& Tripoli's (2024) mixed monomer/TS radius convention on the $72$ 2024 benchmarks likewise gives $95/97$ ($97.9\%$, mean $-2.95\ \mathrm{kcal\ mol^{-1}}$).
Cross-solver differences between PySCF SMD and Jaguar PBF concentrate in strongly polar species: PBF solvates isolated $\mathrm{NH_3}$ more strongly than SMD and experiment ($-6.37\ \mathrm{kcal\ mol^{-1}}$ in PBF vs.\ $-3.31\ \mathrm{kcal\ mol^{-1}}$ in SMD and $-4.29\ \mathrm{kcal\ mol^{-1}}$ experimentally \cite{Marenich2009SMD}) and stabilizes dipolar $\mathrm{S_N2}$ saddles (R91 and R92; Figure~\ref{fig:si_plate7}) by $6.4\text{--}8.2\ \mathrm{kcal\ mol^{-1}}$ more than SMD, shifting best-of-three $\Delta\Delta G^\ddagger$ on R91 from $+0.5$ to $-1.7\ \mathrm{kcal\ mol^{-1}}$ ($+0.9$ to $-0.8\ \mathrm{kcal\ mol^{-1}}$ at $w = 3$, partially offset by stronger $\mathrm{NH_3}$ reactant solvation) and on R92 from $+4.2$ to $-3.1\ \mathrm{kcal\ mol^{-1}}$ ($-1.9\ \mathrm{kcal\ mol^{-1}}$ on the SMD-winning conformer).

Recomputing barriers from the stationary-point tables in Kua et al.\ (2013) and Kua \& Tripoli (2024) identified four transcription or arithmetic errors corrected in our reference table:
(i) a digit transposition in $E_{\mathrm{elec}}$ for \textsc{d}-glyceraldehyde hydration (R9, Figure~\ref{fig:si_plate1}; $-573.00409$ printed for $-573.04010\ \mathrm{Ha}$ in Kua et al.\ 2013 Table~2);
(ii) an omitted minus sign on $-0.5\,TS$ in uncatalyzed C2 Cannizzaro disproportionation (R61, Table~\ref{tab:si_master}; Kua \& Tripoli 2024 Table~S1);
(iii) swapped $\mathrm{NH_3}$-catalyzed C1+C3 aldol rows R76 and R79 (Table~\ref{tab:si_master}; Kua \& Tripoli 2024 Table~S1); and
(iv) a $-5.48\ \mathrm{kcal\ mol^{-1}}$ $\mathrm{NH_3}$ reference offset between figures and Table~S1 in Kua \& Tripoli (2024).

\begin{table*}[!ht]
\centering
\footnotesize
\caption{\textbf{Identical-geometry solvation calibration and agent comparison across published 3D transition states in Kua et al.\ (2013) \cite{Kua2013Formose}.} Kua's 2013 Supporting Information coordinates were re-scored at B3LYP/6-311G** in both Jaguar PBF (\textbf{Kua SI PBF}, \texttt{pbf\_radii=0}) and PySCF SMD (\textbf{Kua SI SMD}, $\mathrm{kcal\ mol^{-1}}$), alongside Kua's printed 2013 barriers (\textbf{Kua 2013 Pub.}). $w_{\mathrm{Kua}}$ and $w^*_{\mathrm{SMD}}$ denote catalytic $\mathrm{H_2O}$ counts (excluding reactant $\mathrm{H_2O}$ in hydrations; \textbf{Agent Best PBF} minimizes across verified active-relay $w$), $\Delta G_{\mathrm{SMD},w_{\mathrm{Kua}}}^\ddagger$ is the best agent SMD barrier at $w = w_{\mathrm{Kua}}$, $\Delta\Delta G_{\mathrm{best,SI}}^\ddagger = \mathrm{Agent\ Best} - \mathrm{Kua\ SI}$ under the matching solver (SMD or PBF), and \textbf{3D} references the side-by-side figure.}
\label{tab:si_kua2013}
\resizebox{\textwidth}{!}{%
\begin{tabular}{@{}llcccccccccccc@{}}
\toprule
\textbf{ID} & \textbf{Reaction} & $w_{\mathrm{Kua}}$ & \shortstack[c]{\textbf{Kua 2013}\\\textbf{Pub.}} & \shortstack[c]{\textbf{Kua SI}\\\textbf{(PBF)}} & \shortstack[c]{\textbf{Kua SI}\\\textbf{(SMD)}} & $\Delta G_{\mathrm{SMD},w_{\mathrm{Kua}}}^\ddagger$ & $w^*_{\mathrm{SMD}}$ & \textbf{Agent Mean $\pm$ SD} & \shortstack[c]{\textbf{Agent Best}\\\textbf{(SMD)}} & \shortstack[c]{$\Delta\Delta G_{\mathrm{best,SI}}^\ddagger$\\\textbf{(SMD)}} & \shortstack[c]{\textbf{Agent Best}\\\textbf{(PBF)}} & \shortstack[c]{$\Delta\Delta G_{\mathrm{best,SI}}^\ddagger$\\\textbf{(PBF)}} & \textbf{3D} \\
\midrule
R8 & $2\,\mathrm{CH_2O} \leftrightarrow$ glycolaldehyde & 2 & 45.29 & 45.46 & 43.65 & 43.81 & 3 & $41.74 \pm 0.97$ & \textbf{40.84} & $-2.81$ & \textbf{42.39} & $-3.07$ & Fig.~\ref{fig:si_plate1} \\
R9 & \textsc{d}-glyceraldehyde + $\mathrm{H_2O}$ hydration & 2 & 14.83 & 16.23 & 16.64 & 14.83 & 3 & $13.88 \pm 0.84$ & \textbf{13.40} & $-3.24$ & \textbf{14.74} & $-1.49$ & Fig.~\ref{fig:si_plate1} \\
R10 & dihydroxyacetone + $\mathrm{H_2O}$ hydration & 2 & 19.90 & 19.25 & 19.55 & 19.09 & 3 & $19.09 \pm 0.30$ & \textbf{18.77} & $-0.78$ & \textbf{18.81} & $-0.44$ & Fig.~\ref{fig:si_plate1} \\
R11 & \textsc{d}-erythrulose + $\mathrm{H_2O}$ hydration & 2 & 19.95 & 17.81 & 17.52 & 17.63 & 3 & $17.30 \pm 0.90$ & \textbf{16.29} & $-1.23$ & \textbf{17.61} & $-0.20$ & Fig.~\ref{fig:si_plate1} \\
R12 & \textsc{d}-threose $\leftrightarrow$ C4-enediol & 3 & 28.69 & 28.71 & 30.10 & 23.06 & 3 & $27.64 \pm 4.54$ & \textbf{23.06} & $-7.04$ & \textbf{22.48} & $-6.23$ & Fig.~\ref{fig:si_plate2} \\
R13 & \textsc{d}-threose + $\mathrm{H_2O}$ hydration & 2 & 15.49 & 15.53 & 16.22 & 13.79 & 3 & $14.15 \pm 1.76$ & \textbf{12.40} & $-3.82$ & \textbf{14.51} & $-1.02$ & Fig.~\ref{fig:si_plate2} \\
R14 & \textsc{d}-threose $\leftrightarrow \beta$-\textsc{d}-threofuranose & 2 & 14.73 & 14.77 & 15.38 & 12.96 & 3 & $12.35 \pm 0.90$ & \textbf{11.57} & $-3.81$ & \textbf{12.66} & $-2.11$ & Fig.~\ref{fig:si_plate2} \\
R15 & \textsc{d}-threose $\leftrightarrow \alpha$-\textsc{d}-threofuranose & 2 & 15.98 & 16.43 & 17.42 & 13.07 & 3 & $13.09 \pm 0.74$ & \textbf{12.37} & $-5.05$ & \textbf{12.74} & $-3.69$ & Fig.~\ref{fig:showcase_3d}, \ref{fig:si_plate2} \\
R16 & \textsc{d}-erythrose $\leftrightarrow \beta$-\textsc{d}-erythrofuranose & 2 & 14.05 & 14.25 & 14.94 & 13.77 & 3 & $15.60 \pm 3.51$ & \textbf{13.38} & $-1.56$ & \textbf{11.86} & $-2.39$ & Fig.~\ref{fig:si_plate3} \\
R17 & \textsc{d}-erythrose $\leftrightarrow \alpha$-\textsc{d}-erythrofuranose & 2 & 11.71 & 11.38 & 11.46 & 11.38 & 3 & $10.59 \pm 1.19$ & \textbf{9.84} & $-1.62$ & \textbf{11.49} & $+0.11$ & Fig.~\ref{fig:si_plate3} \\
R18 & $2\,\mathrm{CH_2O} \rightarrow \mathrm{HCOOCH_3}$ (Tishchenko) & 3 & 61.40 & 64.51 & 61.49 & 57.51 & 0 & $43.00 \pm 0.01$ & \textbf{42.99} & $-18.50$ & \textbf{45.41} & $-19.10$ & Fig.~\ref{fig:showcase_3d}, \ref{fig:si_plate3} \\
R19 & $\mathrm{CH_2O + H_2O} \leftrightarrow \mathrm{CH_2(OH)_2}$ & 2 & 11.57 & 11.84 & 12.45 & 11.96 & 3 & $11.16 \pm 0.01$ & \textbf{11.15} & $-1.30$ & \textbf{11.01} & $-0.83$ & Fig.~\ref{fig:si_plate3} \\
R28 & $\mathrm{CH_2O + CH_2(OH)_2} \rightarrow \mathrm{HCOOH + CH_3OH}$ & 0 & 24.42 & 24.65 & 23.55 & 23.48 & 2 & $23.42 \pm 0.18$ & \textbf{23.22} & $-0.33$ & \textbf{23.16} & $-1.49$ & Fig.~\ref{fig:si_plate4} \\
R37 & glycolaldehyde $\leftrightarrow$ ethenediol & 3 & 21.11 & 24.65 & 25.04 & 23.83 & 4 & $23.01 \pm 0.15$ & \textbf{22.90} & $-2.14$ & \textbf{22.63} & $-2.02$ & Fig.~\ref{fig:si_plate4} \\
R46 & ethenediol + $\mathrm{CH_2O} \leftrightarrow$ \textsc{d}-glyceraldehyde & 2 & 22.31 & 22.26 & 20.87 & 15.26 & 3 & $13.60 \pm 0.52$ & \textbf{13.26} & $-7.61$ & \textbf{15.04} & $-7.22$ & Fig.~\ref{fig:si_plate4} \\
R55 & glycolaldehyde + $\mathrm{H_2O}$ hydration & 2 & 16.05 & 15.99 & 17.03 & 16.27 & 3 & $14.40 \pm 0.49$ & \textbf{14.10} & $-2.93$ & \textbf{15.30} & $-0.69$ & Fig.~\ref{fig:si_plate4} \\
R67 & \textsc{d}-glyceraldehyde $\leftrightarrow$ C3-enediol & 3 & 25.94 & 26.04 & 24.27 & 23.20 & 4 & $24.30 \pm 1.89$ & \textbf{22.15} & $-2.12$ & \textbf{23.53} & $-2.51$ & Fig.~\ref{fig:si_plate5} \\
R70 & dihydroxyacetone $\leftrightarrow$ C3-enediol & 3 & 27.28 & 27.02 & 28.91 & 26.90 & 4 & $26.03 \pm 2.28$ & \textbf{23.44} & $-5.47$ & \textbf{25.28} & $-1.74$ & Fig.~\ref{fig:si_plate5} \\
R74 & C3-enediol + $\mathrm{CH_2O} \leftrightarrow$ \textsc{d}-erythrulose & 2 & 17.63 & 18.01 & 15.36 & 13.29 & 3 & $14.07 \pm 2.50$ & \textbf{12.05} & $-3.31$ & \textbf{14.87} & $-3.14$ & Fig.~\ref{fig:si_plate5} \\
R80 & \textsc{d}-erythrulose $\leftrightarrow$ C4-enediol & 3 & 30.06 & 28.77 & 27.21 & 22.97 & 3 & $23.00 \pm 0.05$ & \textbf{22.97} & $-4.24$ & \textbf{25.85} & $-2.92$ & Fig.~\ref{fig:si_plate5} \\
\bottomrule
\end{tabular}%
}
\end{table*}

\FloatBarrier
\section{Representative Discoveries and Side-by-Side 3D Gallery}
\label{app:gallery}

\begin{figure*}[p]
  \centering
  \includegraphics[width=0.96\textwidth]{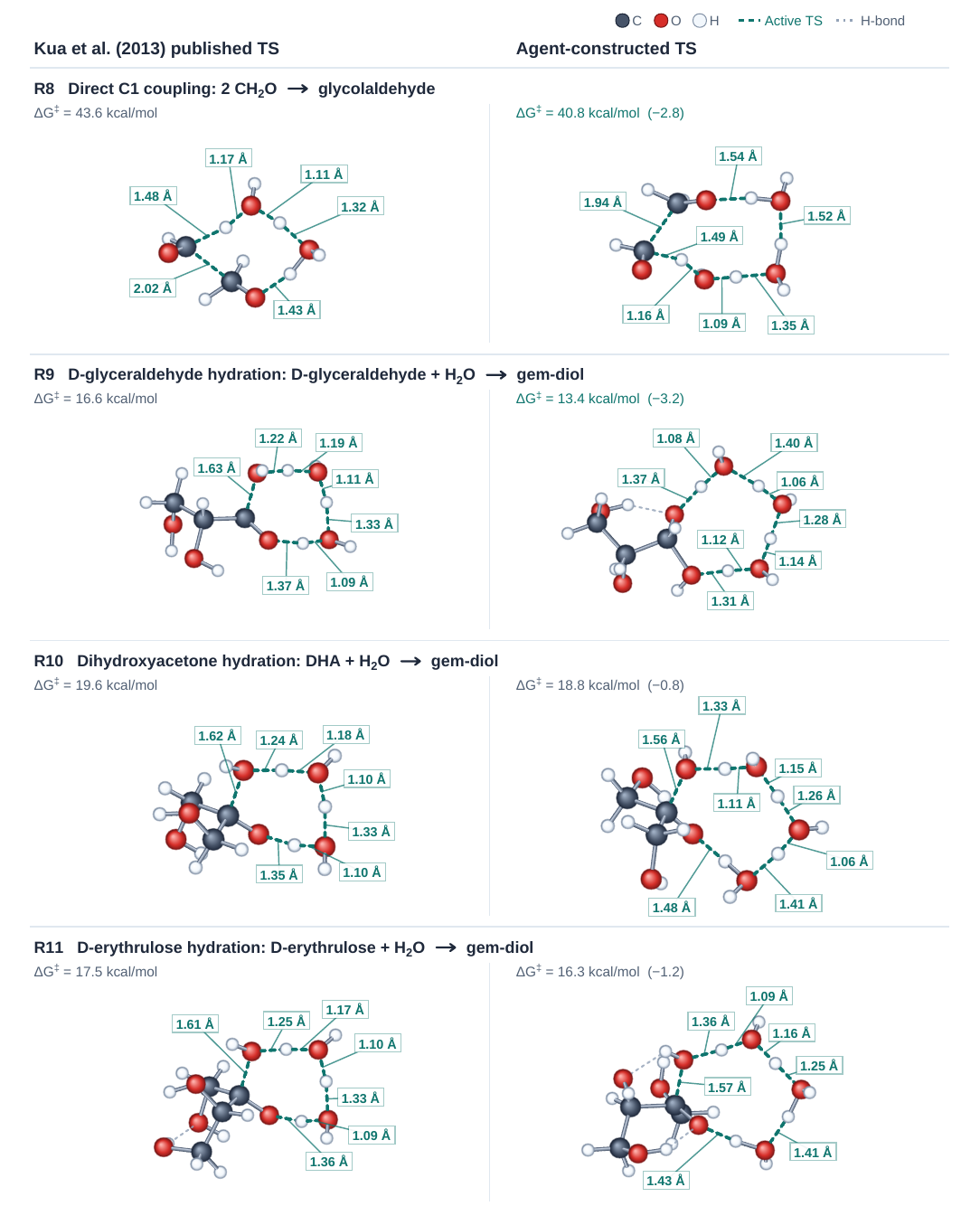}
  \caption{\textbf{Side-by-side 3D transition states (Part 1 of 5: R8--R11).} Left: Kua et al.\ (2013) published geometry. Right: agent-constructed transition state. Displayed barriers $\Delta G^\ddagger$ ($\mathrm{kcal\ mol^{-1}}$) are evaluated identically on both structures at B3LYP/6-311G** ($298\ \mathrm{K}$, PySCF SMD water); parentheses give $\Delta\Delta G^\ddagger = \Delta G_{\mathrm{agent}}^\ddagger - \Delta G_{\mathrm{Kua}}^\ddagger$.}
  \label{fig:si_plate1}
\end{figure*}

\begin{figure*}[p]
  \centering
  \includegraphics[width=0.96\textwidth]{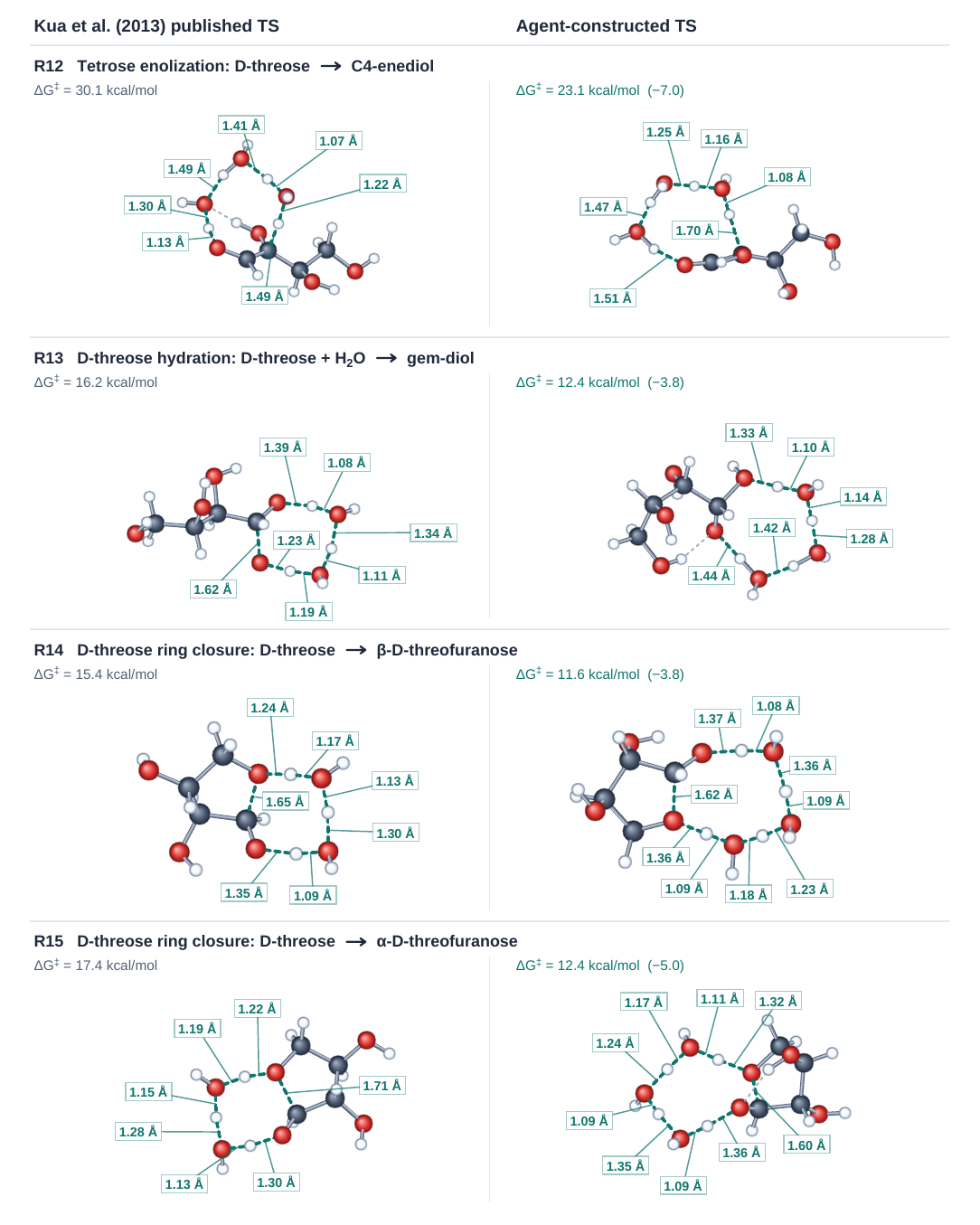}
  \caption{\textbf{Side-by-side 3D transition states (Part 2 of 5: R12--R15).} Left: Kua et al.\ (2013) published geometry. Right: agent-constructed transition state. Same $\Delta G^\ddagger$ ($\mathrm{kcal\ mol^{-1}}$, PySCF SMD) convention as Figure~\ref{fig:si_plate1}.}
  \label{fig:si_plate2}
\end{figure*}

\begin{figure*}[p]
  \centering
  \includegraphics[width=0.96\textwidth]{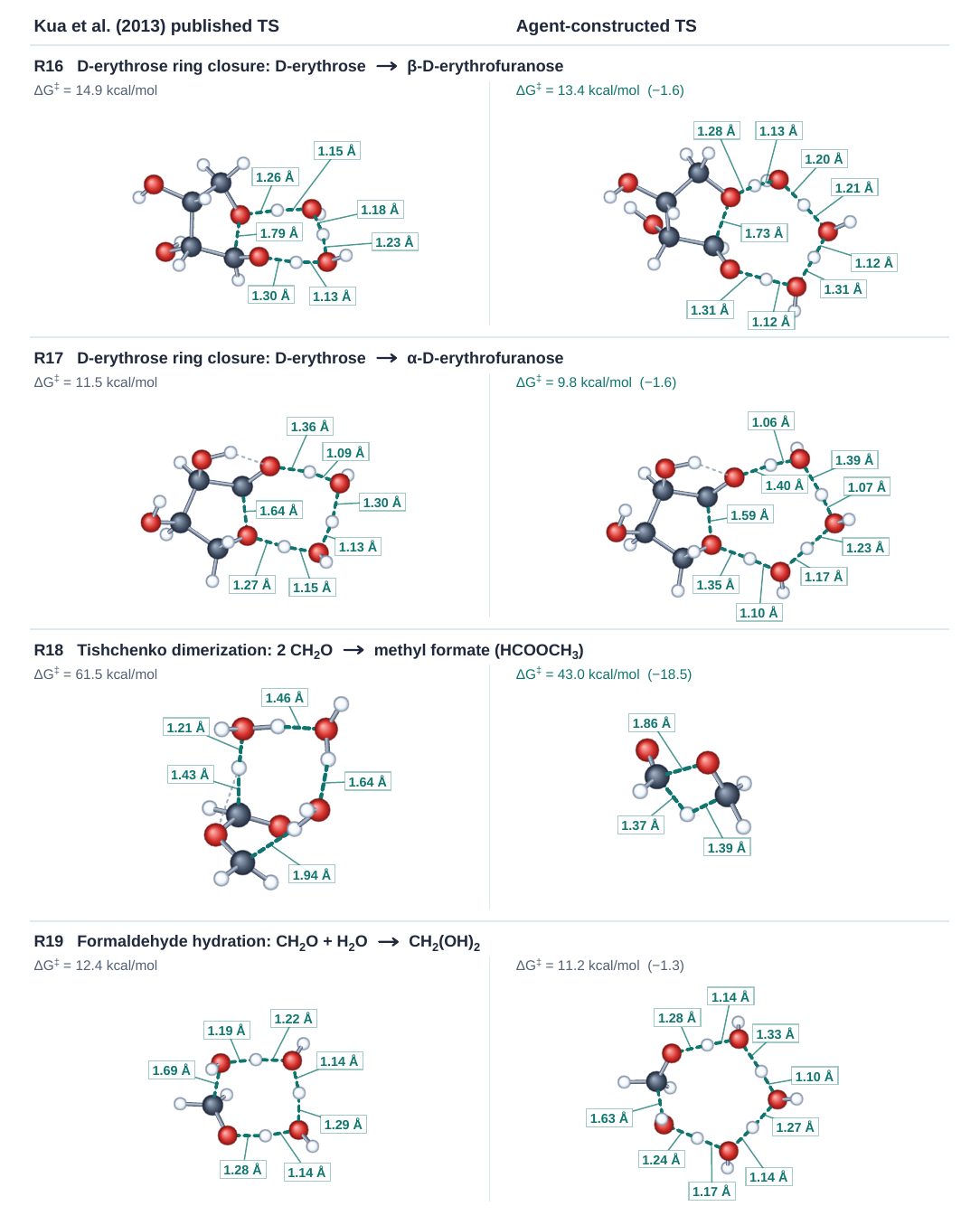}
  \caption{\textbf{Side-by-side 3D transition states (Part 3 of 5: R16--R19).} Left: Kua et al.\ (2013) published geometry. Right: agent-constructed transition state. Same $\Delta G^\ddagger$ ($\mathrm{kcal\ mol^{-1}}$, PySCF SMD) convention as Figure~\ref{fig:si_plate1}.}
  \label{fig:si_plate3}
\end{figure*}

\begin{figure*}[p]
  \centering
  \includegraphics[width=0.96\textwidth]{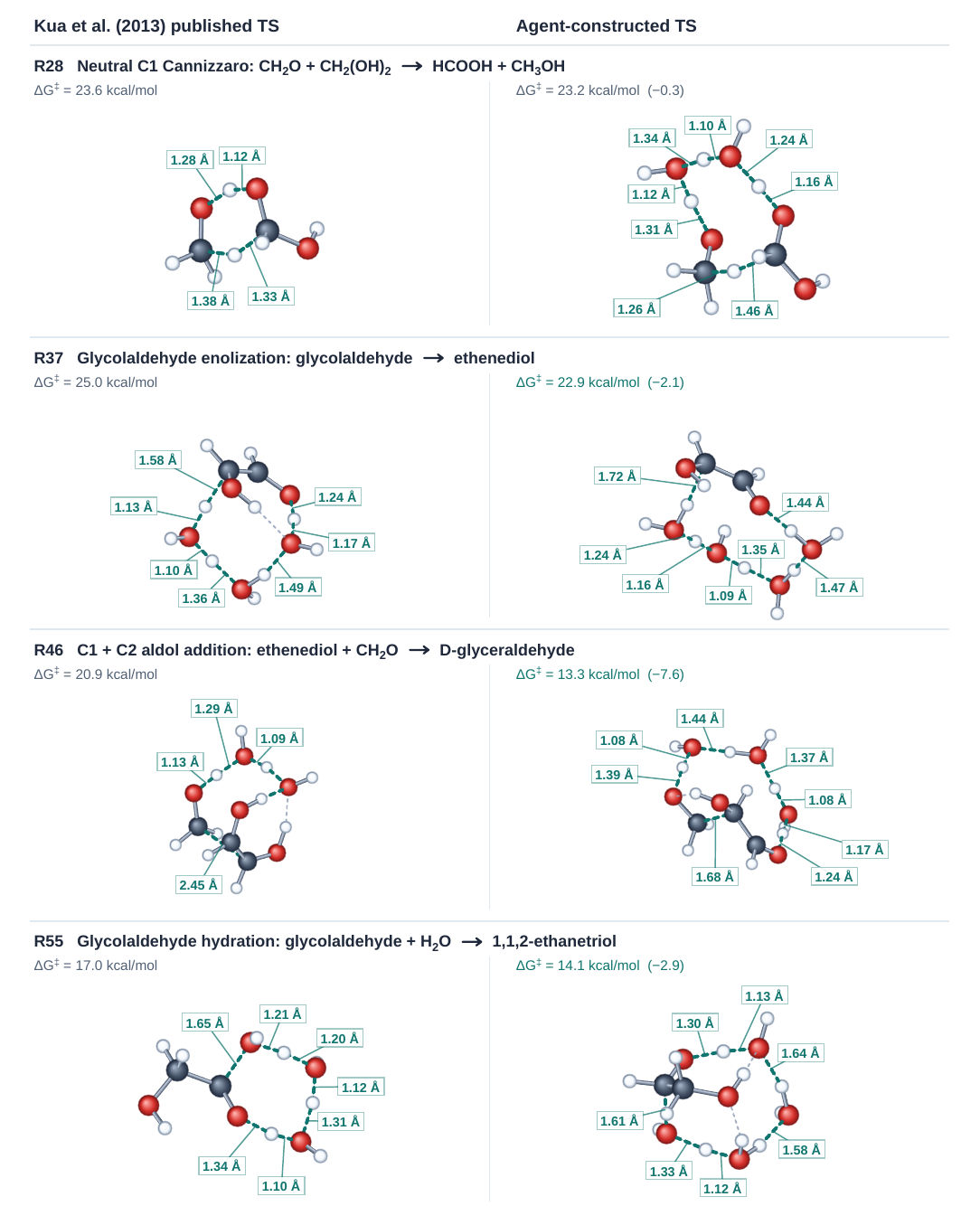}
  \caption{\textbf{Side-by-side 3D transition states (Part 4 of 5: R28, R37, R46, R55).} Left: Kua et al.\ (2013) published geometry. Right: agent-constructed transition state. Same $\Delta G^\ddagger$ ($\mathrm{kcal\ mol^{-1}}$, PySCF SMD) convention as Figure~\ref{fig:si_plate1}.}
  \label{fig:si_plate4}
\end{figure*}

\begin{figure*}[p]
  \centering
  \includegraphics[width=0.96\textwidth]{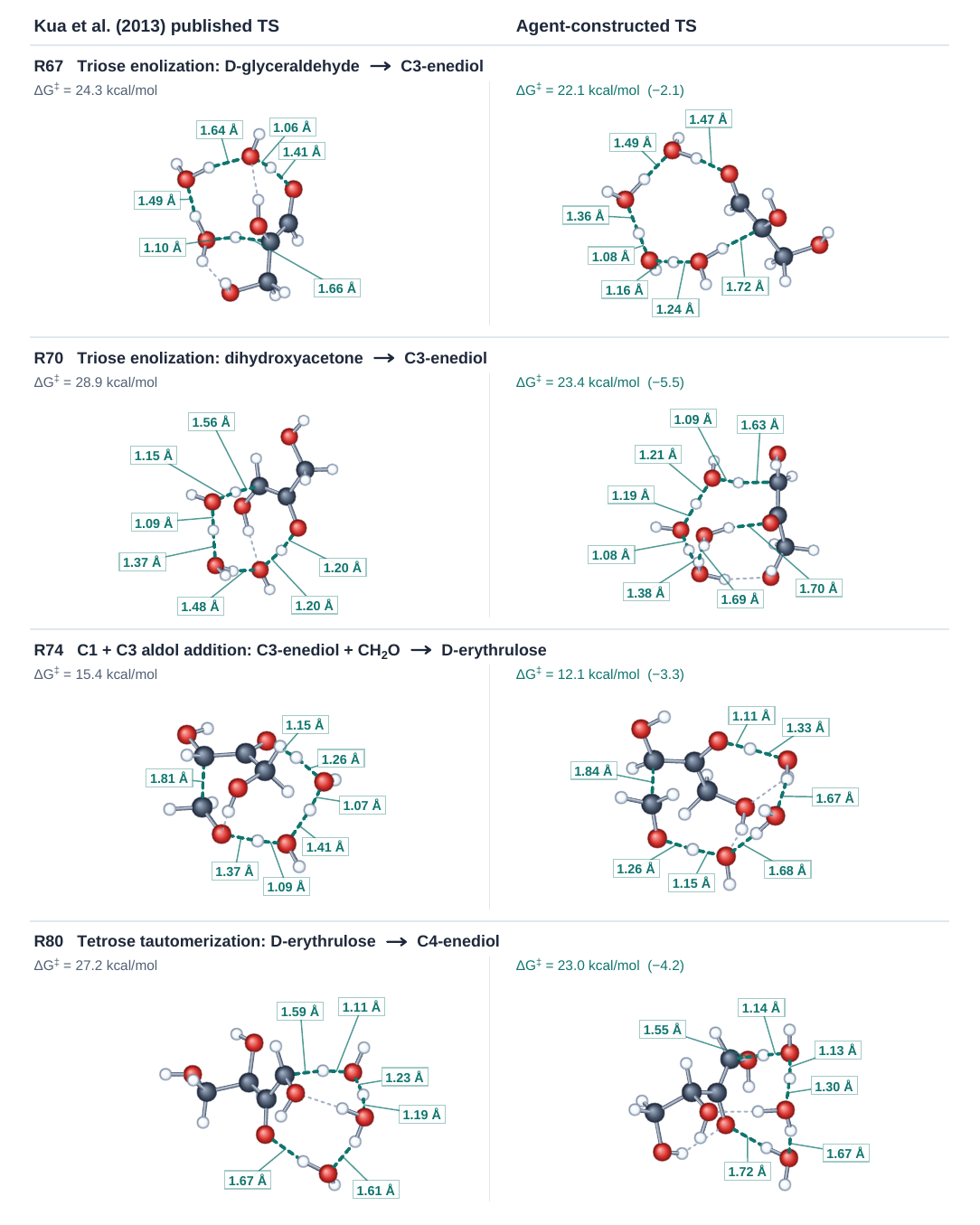}
  \caption{\textbf{Side-by-side 3D transition states (Part 5 of 5: R67, R70, R74, R80).} Left: Kua et al.\ (2013) published geometry. Right: agent-constructed transition state. Same $\Delta G^\ddagger$ ($\mathrm{kcal\ mol^{-1}}$, PySCF SMD) convention as Figure~\ref{fig:si_plate1}.}
  \label{fig:si_plate5}
\end{figure*}

\begin{figure*}[p]
  \centering
  \includegraphics[width=0.96\textwidth]{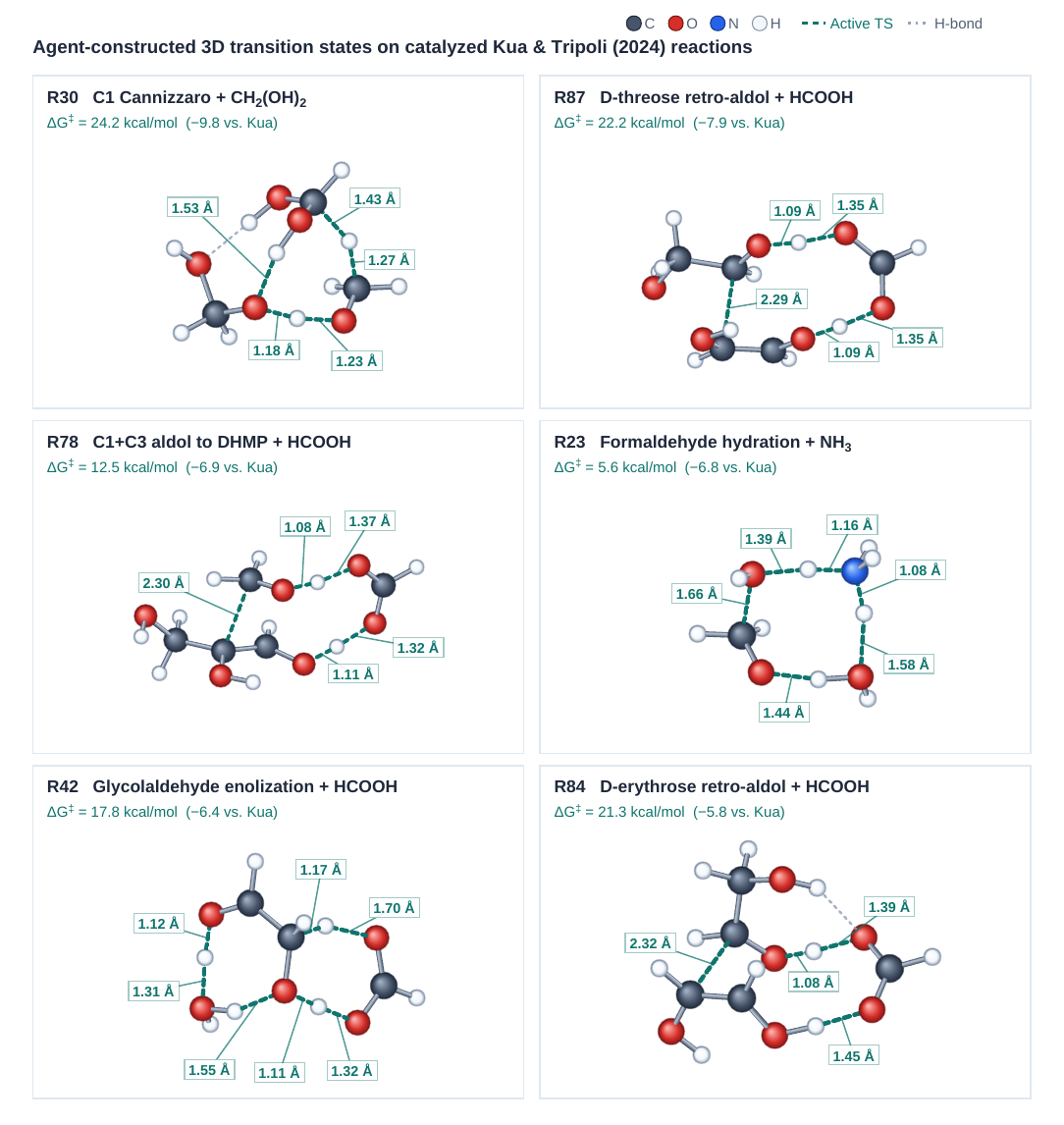}
  \caption{\textbf{Agent-constructed 3D transition states on catalyzed Kua \& Tripoli (2024) reactions (R30, R87, R78, R23, R42, R84).} Shown at Kua's water count ($w = w_{\mathrm{Kua}}$). Displayed barriers $\Delta G^\ddagger$ ($\mathrm{kcal\ mol^{-1}}$) are evaluated at B3LYP/6-311G** ($298\ \mathrm{K}$, PySCF SMD water); because 3D coordinates were not published in Kua \& Tripoli (2024), parenthetical differences ($\Delta\Delta G^\ddagger$) are relative to Kua's published PBF barriers (Table~\ref{tab:si_master}).}
  \label{fig:si_plate6}
\end{figure*}

\begin{figure*}[p]
  \centering
  \includegraphics[width=0.96\textwidth]{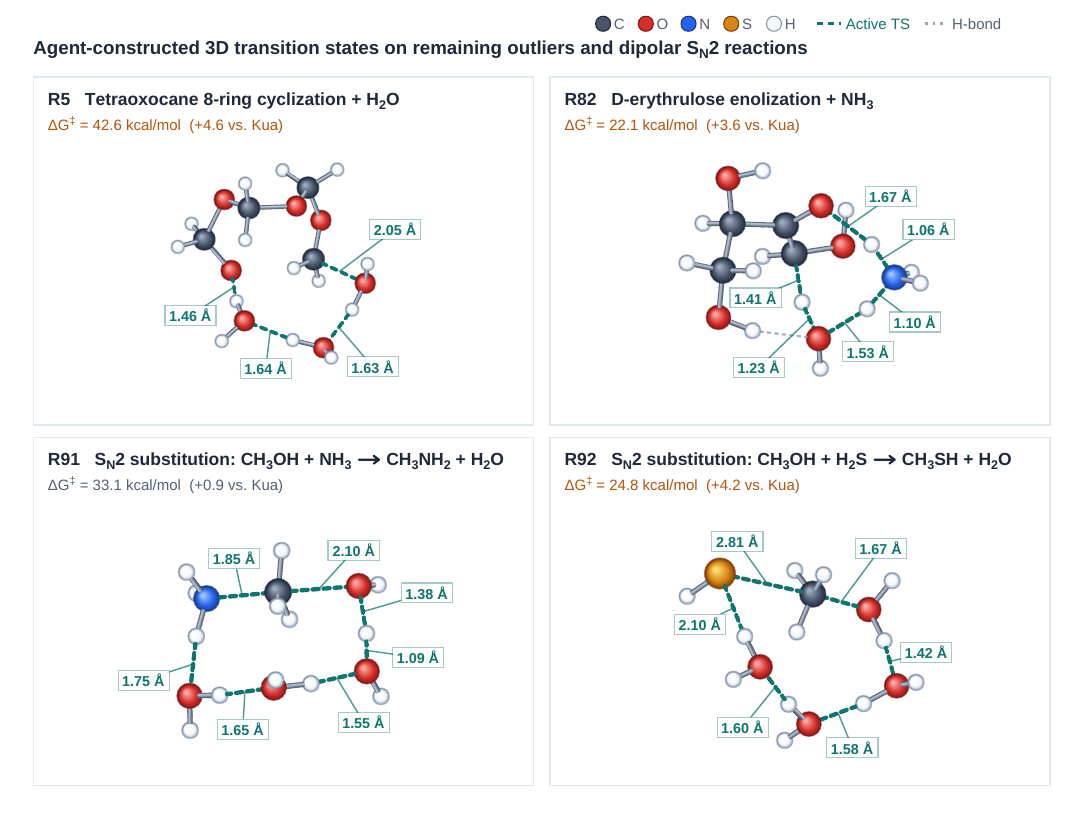}
  \caption{\textbf{Agent-constructed 3D transition states on remaining $> +2\ \mathrm{kcal\ mol^{-1}}$ SMD outliers and dipolar $\mathrm{S_N2}$ reactions (R5, R82, R91, R92).} R5 and R82 show best-of-three geometries ($w^* = 2$ and $1$), whereas R91 and R92 are rendered at matched $w = w_{\mathrm{Kua}} = 3$ to compare dipolar $\mathrm{S_N2}$ solvation directly (R91 reaches $32.7\ \mathrm{kcal\ mol^{-1}}$, $+0.5\ \mathrm{kcal\ mol^{-1}}$, at $w^* = 5$). Displayed barriers $\Delta G^\ddagger$ ($\mathrm{kcal\ mol^{-1}}$) are evaluated at B3LYP/6-311G** ($298\ \mathrm{K}$, PySCF SMD water; parenthetical $\Delta\Delta G^\ddagger$ relative to Kua's published PBF barriers). Under Jaguar PBF re-scoring (Table~\ref{tab:si_master}), R91 and R92 drop to $31.4$ ($-0.8\ \mathrm{kcal\ mol^{-1}}$ at $w=3$; $30.5$, $-1.7\ \mathrm{kcal\ mol^{-1}}$ at $w^*=5$) and $17.6\ \mathrm{kcal\ mol^{-1}}$ ($-3.1\ \mathrm{kcal\ mol^{-1}}$ across replicates; $18.8$, $-1.9\ \mathrm{kcal\ mol^{-1}}$ on the rendered SMD-winning conformer), while R5 and R82 are $40.0$ ($+2.1$) and $23.2\ \mathrm{kcal\ mol^{-1}}$ ($+4.7$).}
  \label{fig:si_plate7}
\end{figure*}

Table~\ref{tab:discoveries} summarizes representative lower-barrier transition states discovered across reaction classes, and Figures~\ref{fig:si_plate1}--\ref{fig:si_plate5} show side-by-side 3D ball-and-stick structures for all $20$ transition states with published Supporting Information coordinates in Kua et al.\ (2013) \cite{Kua2013Formose}.
All displayed barriers $\Delta G^\ddagger$ ($\mathrm{kcal\ mol^{-1}}$) are evaluated at B3LYP/6-311G** ($298\ \mathrm{K}$, PySCF SMD water).
In Figures~\ref{fig:si_plate6}--\ref{fig:si_plate7} (where Kua did not publish 3D coordinates) energy differences in parentheses are relative to Kua's published PBF barriers and thus include some error from switching solvation methods (Table~\ref{tab:si_master}), whereas in Figures~\ref{fig:si_plate1}--\ref{fig:si_plate5} both structures are evaluated with SMD implicit solvation and energies can be compared directly.
Active bond-forming and bond-breaking contacts are drawn as dashed teal lines labeled in \AA; non-reactive intramolecular hydrogen bonds are drawn as dotted grey lines.
Figure~\ref{fig:si_plate6} displays six representative matched-$w$ ($w = w_{\mathrm{Kua}}$) transition states from Kua \& Tripoli (2024) \cite{Kua2024CoreFormose} (R30, R87, R78, R23, R42, and R84), and Figure~\ref{fig:si_plate7} renders the remaining $> +2\ \mathrm{kcal\ mol^{-1}}$ SMD outliers and dipolar $\mathrm{S_N2}$ transition states discussed in Section~\ref{sec:results} and Appendix~\ref{app:solvation} (R5, R82, R91, and R92).

\begin{table*}[!ht]
\centering
\footnotesize
\caption{\textbf{Representative lower-barrier transition states discovered by agents.} All barriers are in $\mathrm{kcal\ mol^{-1}}$ at B3LYP/6-311G** ($298.15\ \mathrm{K}$). Every row below R18 uses Kua's exact catalytic water count ($w = w_{\mathrm{Kua}}$); $\Delta G_{\mathrm{Kua}}^\ddagger$ lists Kua's published PBF barrier (re-scored SI SMD barriers are $61.5$, $20.9$, and $17.4\ \mathrm{kcal\ mol^{-1}}$ on R18, R46, and R15; Table~\ref{tab:si_kua2013}), $\mathrm{Mean \pm SD}$ spans verified replicates at that $w$ ($n/3$), $\Delta E_{\mathrm{elec}}^{\mathrm{TS}} = E_{\mathrm{elec}}^{\mathrm{TS,agent}} - E_{\mathrm{elec}}^{\mathrm{TS,Kua}}$, and \textbf{3D} links to the rendered geometry.}
\label{tab:discoveries}
\resizebox{\textwidth}{!}{%
\begin{tabular}{@{}llccccccll@{}}
\toprule
\textbf{ID} & \textbf{Reaction} & \textbf{Cat.} & $w$ & $\Delta G_{\mathrm{Kua}}^\ddagger$ & \textbf{Best $\Delta G_{\mathrm{SMD}}^\ddagger$} & \textbf{Rep.\ Mean $\pm$ SD} & $\Delta E_{\mathrm{elec}}^{\mathrm{TS}}$ & \textbf{Physical origin of lower barrier} & \textbf{3D} \\
\midrule
R18 & $2\,\mathrm{CH_2O} \rightarrow \mathrm{HCOOCH_3}$ & None & 0 & 61.4 & \textbf{43.0} & $43.0 \pm 0.01$ (3/3) & --- & Direct 1,3-hydride shift ($w=0$) avoids strained 3-water ring & Fig.~\ref{fig:showcase_3d}, \ref{fig:si_plate3} \\
R30 & $\mathrm{CH_2O + CH_2(OH)_2} \rightarrow \mathrm{HCOOH + CH_3OH}$ & $\mathrm{CH_2(OH)_2}$ & 0 & 34.0 & \textbf{24.2} & $24.6 \pm 0.38$ (3/3) & $-6.0$ & Unstrained 8-membered methanediol proton--hydride bridge & Fig.~\ref{fig:si_plate6} \\
R46 & ethenediol + $\mathrm{CH_2O} \rightarrow$ \textsc{d}-glyceraldehyde & $\mathrm{H_2O}$ & 2 & 22.3 & \textbf{15.3} & $15.3 \pm 0.01$ (3/3) & $-1.0$ & $(Z)$-enediol aldol ring with internal H-bond to forming alkoxide & Fig.~\ref{fig:si_plate4} ($w^*=3$) \\
R78 & C3-enediol + $\mathrm{CH_2O} \rightarrow$ DHMP & $\mathrm{HCOOH}$ & 0 & 19.5 & \textbf{12.5} & $13.8 \pm 2.17$ (3/3) & $-4.2$ & Bifunctional $\mathrm{HCOOH}$ bridge concerted with C--C bond formation & Fig.~\ref{fig:si_plate6} \\
R23 & $\mathrm{CH_2O + H_2O} \rightarrow \mathrm{CH_2(OH)_2}$ & $\mathrm{NH_3}$ & 1 & 12.3 & \textbf{5.6} & $5.6 \pm 0.03$ (3/3) & $-3.2$ & 8-membered $\mathrm{NH_3 + H_2O}$ catalytic relay ($+$ reactant $\mathrm{H_2O}$) & Fig.~\ref{fig:si_plate6} \\
R42 & glycolaldehyde $\rightarrow$ ethenediol & $\mathrm{HCOOH}$ & 1 & 24.1 & \textbf{17.8} & $18.7 \pm 0.82$ (3/3) & $-2.7$ & $\alpha$-OH-bridged bicyclic $\mathrm{HCOOH}$/$\mathrm{H_2O}$ relay relieves ring strain & Fig.~\ref{fig:si_plate6} \\
R87 & \textsc{d}-threose $\rightarrow$ ethenediol + glycolaldehyde & $\mathrm{HCOOH}$ & 0 & 30.1 & \textbf{22.2} & $22.5 \pm 0.47$ (3/3) & $-9.8$ & 10-membered carboxylic acid relay across retro-aldol cleavage & Fig.~\ref{fig:si_plate6} \\
R15 & \textsc{d}-threose $\rightarrow \alpha$-\textsc{d}-threofuranose & $\mathrm{H_2O}$ & 2 & 16.0 & \textbf{13.1} & $13.5 \pm 0.54$ (2/3) & $-4.9$ & Envelope sugar pucker preserves an internal hydroxyl H-bond & Fig.~\ref{fig:showcase_3d}, \ref{fig:si_plate2} ($w^*=3$) \\
\bottomrule
\end{tabular}%
}
\end{table*}

\clearpage
\section{Complete 97-Benchmark Comparison Table}
\label{app:master_table}

{\fontsize{5.7}{7.3}\selectfont
\setlength{\tabcolsep}{0.9pt}
\begin{longtable}{@{}llccccccccccccc@{}}
\caption{\textbf{Master comparison across all 97 reactions (R1--R97) in Kua et al.\ (2013) and Kua \& Tripoli (2024) across three independent replicates.} All barriers are in $\mathrm{kcal\ mol^{-1}}$ at $298.15\ \mathrm{K}$. Species shorthand: $\mathrm{1a}$--$\mathrm{4a}$, C1--C4 aldoses ($\mathrm{CH_2O}$, glycolaldehyde, \textsc{d}-glyceraldehyde, \textsc{d}-threose; $\mathrm{4a3}$, \textsc{d}-erythrose; $\mathrm{DHMP}$, 2,3-dihydroxy-2-(hydroxymethyl)propanal); $\mathrm{3k}$/$\mathrm{4k}$, dihydroxyacetone/\textsc{d}-erythrulose; $\mathrm{2e}$--$\mathrm{4e}$, C2--C4 enediols; $\mathrm{1ha}$--$\mathrm{4ha}$ and $\mathrm{3hk}$/$\mathrm{4hk}$, hydrated \emph{gem}-diols ($\mathrm{1ha = CH_2(OH)_2}$); $\mathrm{4f}$/$\mathrm{4f2}$, $\beta$-/$\alpha$-\textsc{d}-threofuranose; $\mathrm{4f3}$/$\mathrm{4f4}$, $\beta$-/$\alpha$-\textsc{d}-erythrofuranose; $\mathrm{2p}$--$\mathrm{4p}$, linear polyoxymethylene diols; $\mathrm{3o}$/$\mathrm{4o}$, 1,3,5-trioxane/1,3,5,7-tetraoxocane; $\mathrm{HOCH_2COOH}$, glycolic acid; $\mathrm{(CH_2OH)_2}$, ethylene glycol. $w_{\mathrm{Kua}}$ and $w^*_{\mathrm{SMD}}$ denote catalytic $\mathrm{H_2O}$ counts (excluding reactant $\mathrm{H_2O}$ in hydrations); \textbf{Cat.} is \texttt{None} when uncatalyzed with $w_{\mathrm{Kua}} = 0$. $n/3$ indicates how many of the 3 replicates made at least one valid submission; $\Delta G_{\mathrm{SMD},w_{\mathrm{Kua}}}^\ddagger$ and $\Delta E_{\mathrm{elec}}^{\mathrm{TS}} = E_{\mathrm{elec}}^{\mathrm{TS,agent}} - E_{\mathrm{elec}}^{\mathrm{TS,Kua}}$ report matched catalytic water count ($w = w_{\mathrm{Kua}}$); $w^*_{\mathrm{SMD}}$, $\mathrm{Mean \pm SD}$ (SMD), $\mathrm{Best\ SMD}$ ($\Delta\Delta G_{\mathrm{SMD}}^\ddagger$), and $\mathrm{Best\ PBF}$ ($\Delta\Delta G_{\mathrm{PBF}}^\ddagger$, Jaguar \texttt{pbf\_radii=0}, minimized across verified active-relay $w$) report the active-relay aqueous barriers; \textbf{3D} links to the rendered geometry.}\label{tab:si_master}\\
\toprule
\textbf{ID} & \textbf{Reaction} & \textbf{Cat.} & $w_{\mathrm{Kua}}$ & $\Delta G_{\mathrm{Kua}}^\ddagger$ & $n/3$ & \shortstack[c]{$\Delta G_{\mathrm{SMD}}^\ddagger$\\$(w_{\mathrm{Kua}})$} & \shortstack[c]{$\Delta E_{\mathrm{elec}}^{\mathrm{TS}}$\\$(w_{\mathrm{Kua}})$} & $w^*_{\mathrm{SMD}}$ & \textbf{Mean $\pm$ SD} & \shortstack[c]{\textbf{Best}\\\textbf{SMD}} & $\Delta\Delta G_{\mathrm{SMD}}^\ddagger$ & \shortstack[c]{\textbf{Best}\\\textbf{PBF}} & $\Delta\Delta G_{\mathrm{PBF}}^\ddagger$ & \textbf{3D} \\
\midrule
\endfirsthead
\toprule
\textbf{ID} & \textbf{Reaction} & \textbf{Cat.} & $w_{\mathrm{Kua}}$ & $\Delta G_{\mathrm{Kua}}^\ddagger$ & $n/3$ & \shortstack[c]{$\Delta G_{\mathrm{SMD}}^\ddagger$\\$(w_{\mathrm{Kua}})$} & \shortstack[c]{$\Delta E_{\mathrm{elec}}^{\mathrm{TS}}$\\$(w_{\mathrm{Kua}})$} & $w^*_{\mathrm{SMD}}$ & \textbf{Mean $\pm$ SD} & \shortstack[c]{\textbf{Best}\\\textbf{SMD}} & $\Delta\Delta G_{\mathrm{SMD}}^\ddagger$ & \shortstack[c]{\textbf{Best}\\\textbf{PBF}} & $\Delta\Delta G_{\mathrm{PBF}}^\ddagger$ & \textbf{3D} \\
\midrule
\endhead
\bottomrule
\endfoot
R1 & $\mathrm{1ha} + \mathrm{CH_2O} \,{\to}\, \mathrm{2p}$ & None & 2 & 10.7 & 3/3 & 11.0 & $+1.3$ & 3 & $18.8 \pm 16.0$ & \textbf{9.5} & $-1.2$ & \textbf{11.2} & $+0.5$ & --- \\
R2 & $\mathrm{2p} + \mathrm{CH_2O} \,{\to}\, \mathrm{3p}$ & None & 2 & 12.6 & 3/3 & 10.3 & $-1.0$ & 3 & $10.1 \pm 1.8$ & \textbf{8.0} & $-4.6$ & \textbf{9.3} & $-3.2$ & --- \\
R3 & $\mathrm{3p} + \mathrm{CH_2O} \,{\to}\, \mathrm{4p}$ & None & 2 & 15.1 & 3/3 & 10.4 & $-4.3$ & 3 & $10.4 \pm 2.6$ & \textbf{7.7} & $-7.4$ & \textbf{11.1} & $-4.0$ & --- \\
R4 & $\mathrm{3p} \,{\to}\, \mathrm{3o} + \mathrm{H_2O}$ & None & 3 & 34.6 & 2/3 & 35.5 & $-0.1$ & 3 & $35.5 \pm 0.0$ & \textbf{35.5} & $+0.9$ & \textbf{36.4} & $+1.9$ & --- \\
R5 & $\mathrm{4p} \,{\to}\, \mathrm{4o} + \mathrm{H_2O}$ & None & 3 & 38.0 & 3/3 & --- & --- & 2 & $45.3 \pm 4.8$ & \textbf{42.6} & $+4.6$ & \textbf{40.0} & $+2.1$ & Fig.~\ref{fig:si_plate7} \\
R6 & $\mathrm{4p} \,{\to}\, \mathrm{3o} + \mathrm{1ha}$ & None & 3 & 35.6 & 2/3 & 34.9 & $-0.9$ & 3 & $35.9 \pm 1.4$ & \textbf{34.9} & $-0.7$ & \textbf{35.6} & $0.0$ & --- \\
R7 & $3\,\mathrm{CH_2O} \,{\to}\, \mathrm{3o}$ & None & 0 & 24.4 & 3/3 & 24.2 & $-0.1$ & 0 & $24.2 \pm 0.0$ & \textbf{24.2} & $-0.2$ & \textbf{24.6} & $+0.3$ & --- \\
R8 & $2\,\mathrm{CH_2O} \,{\to}\, \mathrm{2a}$ & None & 2 & 45.3 & 3/3 & 43.8 & $-0.1$ & 3 & $41.7 \pm 1.0$ & \textbf{40.8} & $-4.5$ & \textbf{42.4} & $-2.9$ & Fig.~\ref{fig:si_plate1} \\
R9 & $\mathrm{3a} + \mathrm{H_2O} \,{\to}\, \mathrm{3ha}$ & None & 2 & 14.8 & 3/3 & 14.8 & $-5.3$ & 3 & $13.9 \pm 0.8$ & \textbf{13.4} & $-1.4$ & \textbf{14.7} & $-0.1$ & Fig.~\ref{fig:si_plate1} \\
R10 & $\mathrm{3k} + \mathrm{H_2O} \,{\to}\, \mathrm{3hk}$ & None & 2 & 19.9 & 3/3 & 19.1 & $-1.6$ & 3 & $19.1 \pm 0.3$ & \textbf{18.8} & $-1.1$ & \textbf{18.8} & $-1.1$ & Fig.~\ref{fig:si_plate1} \\
R11 & $\mathrm{4k} + \mathrm{H_2O} \,{\to}\, \mathrm{4hk}$ & None & 2 & 19.9 & 3/3 & 17.6 & $+0.4$ & 3 & $17.3 \pm 0.9$ & \textbf{16.3} & $-3.7$ & \textbf{17.6} & $-2.3$ & Fig.~\ref{fig:si_plate1} \\
R12 & $\mathrm{4e} \,{\to}\, \mathrm{4a}$ & None & 3 & 28.7 & 3/3 & 23.1 & $-0.8$ & 3 & $27.6 \pm 4.5$ & \textbf{23.1} & $-5.6$ & \textbf{22.5} & $-6.2$ & Fig.~\ref{fig:si_plate2} \\
R13 & $\mathrm{4a} + \mathrm{H_2O} \,{\to}\, \mathrm{4ha}$ & None & 2 & 15.5 & 3/3 & 13.8 & $-7.2$ & 3 & $14.2 \pm 1.8$ & \textbf{12.4} & $-3.1$ & \textbf{14.5} & $-1.0$ & Fig.~\ref{fig:si_plate2} \\
R14 & $\mathrm{4a} \,{\to}\, \mathrm{4f}$ & None & 2 & 14.7 & 3/3 & 13.0 & $-3.7$ & 3 & $12.3 \pm 0.9$ & \textbf{11.6} & $-3.2$ & \textbf{12.7} & $-2.1$ & Fig.~\ref{fig:si_plate2} \\
R15 & $\mathrm{4a} \,{\to}\, \mathrm{4f2}$ & None & 2 & 16.0 & 3/3 & 13.1 & $-4.9$ & 3 & $13.1 \pm 0.7$ & \textbf{12.4} & $-3.6$ & \textbf{12.7} & $-3.2$ & Fig.~\ref{fig:showcase_3d}, \ref{fig:si_plate2} \\
R16 & $\mathrm{4a3} \,{\to}\, \mathrm{4f3}$ & None & 2 & 14.1 & 3/3 & 13.8 & $+2.0$ & 3 & $15.6 \pm 3.5$ & \textbf{13.4} & $-0.7$ & \textbf{11.9} & $-2.2$ & Fig.~\ref{fig:si_plate3} \\
R17 & $\mathrm{4a3} \,{\to}\, \mathrm{4f4}$ & None & 2 & 11.7 & 3/3 & 11.4 & $-0.1$ & 3 & $10.6 \pm 1.2$ & \textbf{9.8} & $-1.9$ & \textbf{11.5} & $-0.2$ & Fig.~\ref{fig:si_plate3} \\
R18 & $2\,\mathrm{CH_2O} \,{\to}\, \mathrm{HCOOCH_3}$ & None & 3 & 61.4 & 3/3 & 57.5 & $-7.5$ & 0 & $43.0 \pm 0.0$ & \textbf{43.0} & $-18.4$ & \textbf{45.4} & $-16.0$ & Fig.~\ref{fig:showcase_3d}, \ref{fig:si_plate3} \\
R19 & $\mathrm{CH_2O} + \mathrm{H_2O} \,{\to}\, \mathrm{1ha}$ & None & 2 & 11.6 & 3/3 & 12.0 & $+0.8$ & 3 & $11.2 \pm 0.0$ & \textbf{11.2} & $-0.4$ & \textbf{11.0} & $-0.6$ & Fig.~\ref{fig:si_plate3} \\
R20 & $\mathrm{CH_2O} + \mathrm{H_2O} \,{\to}\, \mathrm{1ha}$ & $\mathrm{CH_3OH}$ & 1 & 13.7 & 3/3 & 11.2 & $+0.4$ & 2 & $10.5 \pm 0.6$ & \textbf{10.1} & $-3.5$ & \textbf{11.7} & $-1.9$ & --- \\
R21 & $\mathrm{CH_2O} + \mathrm{H_2O} \,{\to}\, \mathrm{1ha}$ & $\mathrm{1ha}$ & 1 & 16.2 & 3/3 & 10.2 & $-2.4$ & 2 & $8.7 \pm 0.4$ & \textbf{8.3} & $-7.9$ & \textbf{10.8} & $-5.4$ & --- \\
R22 & $\mathrm{CH_2O} + \mathrm{H_2O} \,{\to}\, \mathrm{1ha}$ & $\mathrm{H_2S}$ & 1 & 12.4 & 3/3 & 13.4 & $+0.4$ & 2 & $12.8 \pm 0.6$ & \textbf{12.2} & $-0.1$ & \textbf{9.1} & $-3.3$ & --- \\
R23 & $\mathrm{CH_2O} + \mathrm{H_2O} \,{\to}\, \mathrm{1ha}$ & $\mathrm{NH_3}$ & 1 & 12.3 & 3/3 & 5.6 & $-3.2$ & 1 & $5.6 \pm 0.0$ & \textbf{5.6} & $-6.8$ & \textbf{6.9} & $-5.4$ & Fig.~\ref{fig:si_plate6} \\
R24 & $\mathrm{CH_2O} + \mathrm{H_2O} \,{\to}\, \mathrm{1ha}$ & $\mathrm{HCOOH}$ & 0 & 7.8 & 3/3 & 5.2 & $-2.6$ & 1 & $4.0 \pm 0.0$ & \textbf{4.0} & $-3.8$ & \textbf{4.3} & $-3.5$ & --- \\
R25 & $\mathrm{CH_2O} + \mathrm{H_2O} \,{\to}\, \mathrm{1ha}$ & $\mathrm{H_2CO_3}$ & 0 & 6.1 & 3/3 & 3.9 & $-2.6$ & 1 & $2.5 \pm 0.1$ & \textbf{2.4} & $-3.8$ & \textbf{3.0} & $-3.1$ & --- \\
R26 & $\mathrm{CH_2O} + \mathrm{H_2O} \,{\to}\, \mathrm{1ha}$ & $\mathrm{CH_3COOH}$ & 1 & 4.6 & 3/3 & 5.0 & $-0.1$ & 1 & $5.2 \pm 0.5$ & \textbf{5.0} & $+0.3$ & \textbf{5.6} & $+1.0$ & --- \\
R27 & $\mathrm{CH_2O} + \mathrm{H_2O} \,{\to}\, \mathrm{1ha}$ & $\mathrm{HOCH_2COOH}$ & 1 & 6.5 & 3/3 & 3.4 & $-6.2$ & 1 & $3.4 \pm 0.0$ & \textbf{3.4} & $-3.0$ & \textbf{3.9} & $-2.6$ & --- \\
R28 & $\mathrm{CH_2O} + \mathrm{1ha} \,{\to}\, \mathrm{HCOOH} + \mathrm{CH_3OH}$ & None & 0 & 24.4 & 3/3 & 23.5 & $-0.1$ & 2 & $23.4 \pm 0.2$ & \textbf{23.2} & $-1.2$ & \textbf{23.2} & $-1.3$ & Fig.~\ref{fig:si_plate4} \\
R29 & $\mathrm{CH_2O} + \mathrm{1ha} \,{\to}\, \mathrm{HCOOH} + \mathrm{CH_3OH}$ & $\mathrm{CH_3OH}$ & 0 & 30.3 & 3/3 & 26.3 & $-1.6$ & 3 & $23.4 \pm 0.7$ & \textbf{22.6} & $-7.7$ & \textbf{23.4} & $-7.0$ & --- \\
R30 & $\mathrm{CH_2O} + \mathrm{1ha} \,{\to}\, \mathrm{HCOOH} + \mathrm{CH_3OH}$ & $\mathrm{1ha}$ & 0 & 34.0 & 3/3 & 24.2 & $-6.0$ & 1 & $22.4 \pm 2.1$ & \textbf{20.1} & $-13.9$ & \textbf{23.6} & $-10.4$ & Fig.~\ref{fig:si_plate6} \\
R31 & $\mathrm{CH_2O} + \mathrm{1ha} \,{\to}\, \mathrm{HCOOH} + \mathrm{CH_3OH}$ & $\mathrm{H_2S}$ & 0 & 32.4 & 3/3 & 31.5 & $+0.1$ & 3 & $25.6 \pm 0.7$ & \textbf{24.9} & $-7.4$ & \textbf{22.5} & $-9.8$ & --- \\
R32 & $\mathrm{CH_2O} + \mathrm{1ha} \,{\to}\, \mathrm{HCOOH} + \mathrm{CH_3OH}$ & $\mathrm{NH_3}$ & 0 & 23.7 & 3/3 & 22.0 & $-1.2$ & 1 & $19.9 \pm 1.8$ & \textbf{18.7} & $-5.0$ & \textbf{19.1} & $-4.6$ & --- \\
R33 & $\mathrm{CH_2O} + \mathrm{1ha} \,{\to}\, \mathrm{HCOOH} + \mathrm{CH_3OH}$ & $\mathrm{HCOOH}$ & 0 & 20.1 & 3/3 & 17.4 & $0.0$ & 2 & $17.2 \pm 0.5$ & \textbf{16.5} & $-3.6$ & \textbf{17.5} & $-2.6$ & --- \\
R34 & $\mathrm{CH_2O} + \mathrm{1ha} \,{\to}\, \mathrm{HCOOH} + \mathrm{CH_3OH}$ & $\mathrm{H_2CO_3}$ & 0 & 18.6 & 3/3 & 16.1 & $0.0$ & 0 & $16.2 \pm 0.1$ & \textbf{16.1} & $-2.5$ & \textbf{16.9} & $-1.7$ & --- \\
R35 & $\mathrm{CH_2O} + \mathrm{1ha} \,{\to}\, \mathrm{HCOOH} + \mathrm{CH_3OH}$ & $\mathrm{CH_3COOH}$ & 0 & 18.9 & 3/3 & 17.8 & $0.0$ & 0 & $17.9 \pm 0.1$ & \textbf{17.8} & $-1.1$ & \textbf{19.5} & $+0.6$ & --- \\
R36 & $\mathrm{CH_2O} + \mathrm{1ha} \,{\to}\, \mathrm{HCOOH} + \mathrm{CH_3OH}$ & $\mathrm{HOCH_2COOH}$ & 0 & 22.2 & 3/3 & 17.1 & $-5.5$ & 0 & $17.2 \pm 0.3$ & \textbf{17.1} & $-5.2$ & \textbf{18.9} & $-3.4$ & --- \\
R37 & $\mathrm{2a} \,{\to}\, \mathrm{2e}$ & None & 3 & 21.1 & 3/3 & 23.8 & $+2.7$ & 4 & $23.0 \pm 0.1$ & \textbf{22.9} & $+1.8$ & \textbf{22.6} & $+1.5$ & Fig.~\ref{fig:si_plate4} \\
R38 & $\mathrm{2a} \,{\to}\, \mathrm{2e}$ & $\mathrm{CH_3OH}$ & 2 & 25.2 & 3/3 & 23.0 & $-1.4$ & 3 & $22.5 \pm 1.1$ & \textbf{21.9} & $-3.3$ & \textbf{22.1} & $-3.0$ & --- \\
R39 & $\mathrm{2a} \,{\to}\, \mathrm{2e}$ & $\mathrm{1ha}$ & 2 & 28.8 & 3/3 & 22.1 & $-8.2$ & 3 & $22.5 \pm 1.3$ & \textbf{21.5} & $-7.3$ & \textbf{23.5} & $-5.3$ & --- \\
R40 & $\mathrm{2a} \,{\to}\, \mathrm{2e}$ & $\mathrm{H_2S}$ & 2 & 28.4 & 3/3 & 23.6 & $-6.1$ & 2 & $27.0 \pm 2.9$ & \textbf{23.6} & $-4.8$ & \textbf{18.8} & $-9.5$ & --- \\
R41 & $\mathrm{2a} \,{\to}\, \mathrm{2e}$ & $\mathrm{NH_3}$ & 2 & 17.7 & 3/3 & 19.4 & $+7.2$ & 1 & $19.9 \pm 2.0$ & \textbf{18.1} & $+0.4$ & \textbf{17.9} & $+0.2$ & --- \\
R42 & $\mathrm{2a} \,{\to}\, \mathrm{2e}$ & $\mathrm{HCOOH}$ & 1 & 24.1 & 3/3 & 17.8 & $-2.7$ & 3 & $16.6 \pm 0.3$ & \textbf{16.2} & $-7.9$ & \textbf{16.6} & $-7.5$ & Fig.~\ref{fig:si_plate6} \\
R43 & $\mathrm{2a} \,{\to}\, \mathrm{2e}$ & $\mathrm{H_2CO_3}$ & 1 & 22.5 & 3/3 & 18.2 & $-6.6$ & 1 & $20.5 \pm 3.9$ & \textbf{18.2} & $-4.3$ & \textbf{19.5} & $-3.0$ & --- \\
R44 & $\mathrm{2a} \,{\to}\, \mathrm{2e}$ & $\mathrm{CH_3COOH}$ & 1 & 22.8 & 3/3 & 20.4 & $-5.9$ & 2 & $20.2 \pm 0.2$ & \textbf{20.0} & $-2.8$ & \textbf{20.6} & $-2.2$ & --- \\
R45 & $\mathrm{2a} \,{\to}\, \mathrm{2e}$ & $\mathrm{HOCH_2COOH}$ & 1 & 24.9 & 3/3 & 19.0 & $-11.7$ & 3 & $18.4 \pm 2.3$ & \textbf{15.9} & $-8.9$ & \textbf{17.1} & $-7.8$ & --- \\
R46 & $\mathrm{2e} + \mathrm{CH_2O} \,{\to}\, \mathrm{3a}$ & None & 2 & 22.3 & 3/3 & 15.3 & $-1.0$ & 3 & $13.6 \pm 0.5$ & \textbf{13.3} & $-9.1$ & \textbf{15.0} & $-7.3$ & Fig.~\ref{fig:si_plate4} \\
R47 & $\mathrm{2e} + \mathrm{CH_2O} \,{\to}\, \mathrm{3a}$ & $\mathrm{CH_3OH}$ & 0 & 24.8 & 3/3 & 19.6 & $-0.1$ & 1 & $18.1 \pm 5.3$ & \textbf{15.0} & $-9.8$ & \textbf{19.5} & $-5.3$ & --- \\
R48 & $\mathrm{2e} + \mathrm{CH_2O} \,{\to}\, \mathrm{3a}$ & $\mathrm{1ha}$ & 0 & 29.4 & 3/3 & 21.7 & $+1.5$ & 1 & $17.5 \pm 2.7$ & \textbf{15.5} & $-13.9$ & \textbf{21.5} & $-7.9$ & --- \\
R49 & $\mathrm{2e} + \mathrm{CH_2O} \,{\to}\, \mathrm{3a}$ & $\mathrm{H_2S}$ & 0 & 32.9 & 3/3 & 27.6 & $-1.8$ & 1 & $25.2 \pm 4.4$ & \textbf{22.3} & $-10.6$ & \textbf{22.9} & $-10.0$ & --- \\
R50 & $\mathrm{2e} + \mathrm{CH_2O} \,{\to}\, \mathrm{3a}$ & $\mathrm{NH_3}$ & 0 & 15.5 & 3/3 & 12.9 & $0.0$ & 1 & $11.3 \pm 1.3$ & \textbf{9.8} & $-5.7$ & \textbf{12.1} & $-3.5$ & --- \\
R51 & $\mathrm{2e} + \mathrm{CH_2O} \,{\to}\, \mathrm{3a}$ & $\mathrm{HCOOH}$ & 0 & 17.8 & 3/3 & 13.2 & $-2.5$ & 2 & $12.0 \pm 0.8$ & \textbf{11.1} & $-6.7$ & \textbf{13.5} & $-4.3$ & --- \\
R52 & $\mathrm{2e} + \mathrm{CH_2O} \,{\to}\, \mathrm{3a}$ & $\mathrm{H_2CO_3}$ & 0 & 16.2 & 3/3 & 11.9 & $-2.2$ & 1 & $11.2 \pm 0.6$ & \textbf{10.9} & $-5.4$ & \textbf{13.2} & $-3.0$ & --- \\
R53 & $\mathrm{2e} + \mathrm{CH_2O} \,{\to}\, \mathrm{3a}$ & $\mathrm{CH_3COOH}$ & 0 & 15.6 & 3/3 & 14.6 & $-2.6$ & 1 & $14.2 \pm 0.5$ & \textbf{13.6} & $-2.0$ & \textbf{16.4} & $+0.7$ & --- \\
R54 & $\mathrm{2e} + \mathrm{CH_2O} \,{\to}\, \mathrm{3a}$ & $\mathrm{HOCH_2COOH}$ & 0 & 19.8 & 3/3 & 13.5 & $-8.2$ & 3 & $11.3 \pm 1.0$ & \textbf{10.2} & $-9.6$ & \textbf{14.4} & $-5.4$ & --- \\
R55 & $\mathrm{2a} + \mathrm{H_2O} \,{\to}\, \mathrm{2ha}$ & None & 2 & 16.1 & 3/3 & 16.3 & $-2.9$ & 3 & $14.4 \pm 0.5$ & \textbf{14.1} & $-1.9$ & \textbf{15.3} & $-0.8$ & Fig.~\ref{fig:si_plate4} \\
R56 & $\mathrm{2a} + \mathrm{H_2O} \,{\to}\, \mathrm{2ha}$ & $\mathrm{HCOOH}$ & 1 & 11.0 & 3/3 & 8.6 & $-3.4$ & 2 & $8.5 \pm 0.2$ & \textbf{8.2} & $-2.8$ & \textbf{8.3} & $-2.7$ & --- \\
R57 & $\mathrm{2a} + \mathrm{H_2O} \,{\to}\, \mathrm{2ha}$ & $\mathrm{NH_3}$ & 1 & 15.6 & 3/3 & 8.9 & $-2.2$ & 1 & $9.2 \pm 0.3$ & \textbf{8.9} & $-6.6$ & \textbf{7.8} & $-7.8$ & --- \\
R58 & $\mathrm{2ha} + \mathrm{2a} \,{\to}\, \mathrm{HOCH_2COOH} + \mathrm{(CH_2OH)_2}$ & None & 0 & 30.0 & 3/3 & 26.0 & $+1.4$ & 3 & $25.6 \pm 1.4$ & \textbf{24.4} & $-5.6$ & \textbf{25.0} & $-5.0$ & --- \\
R59 & $\mathrm{2ha} + \mathrm{2a} \,{\to}\, \mathrm{HOCH_2COOH} + \mathrm{(CH_2OH)_2}$ & $\mathrm{HCOOH}$ & 0 & 26.8 & 3/3 & 20.7 & $-0.6$ & 0 & $20.9 \pm 0.3$ & \textbf{20.7} & $-6.1$ & \textbf{23.6} & $-3.2$ & --- \\
R60 & $\mathrm{2ha} + \mathrm{2a} \,{\to}\, \mathrm{HOCH_2COOH} + \mathrm{(CH_2OH)_2}$ & $\mathrm{NH_3}$ & 0 & 28.8 & 3/3 & 23.3 & $+1.3$ & 2 & $21.7 \pm 3.4$ & \textbf{18.2} & $-10.5$ & \textbf{20.4} & $-8.4$ & --- \\
R61 & $\mathrm{2ha} + \mathrm{CH_2O} \,{\to}\, \mathrm{HOCH_2COOH} + \mathrm{CH_3OH}$ & None & 0 & 26.0 & 3/3 & 23.3 & $-1.1$ & 3 & $22.8 \pm 0.8$ & \textbf{21.9} & $-4.1$ & \textbf{21.5} & $-4.5$ & --- \\
R62 & $\mathrm{2ha} + \mathrm{CH_2O} \,{\to}\, \mathrm{HOCH_2COOH} + \mathrm{CH_3OH}$ & $\mathrm{HCOOH}$ & 0 & 21.6 & 2/3 & 17.3 & $-0.1$ & 0 & $17.3 \pm 0.0$ & \textbf{17.3} & $-4.3$ & \textbf{18.6} & $-3.0$ & --- \\
R63 & $\mathrm{2ha} + \mathrm{CH_2O} \,{\to}\, \mathrm{HOCH_2COOH} + \mathrm{CH_3OH}$ & $\mathrm{NH_3}$ & 0 & 24.3 & 3/3 & 20.3 & $-3.7$ & 1 & $17.5 \pm 0.1$ & \textbf{17.4} & $-6.8$ & \textbf{17.5} & $-6.8$ & --- \\
R64 & $\mathrm{2a} + \mathrm{1ha} \,{\to}\, \mathrm{HCOOH} + \mathrm{(CH_2OH)_2}$ & None & 0 & 30.6 & 3/3 & 27.5 & $-3.7$ & 2 & $27.4 \pm 0.2$ & \textbf{27.1} & $-3.4$ & \textbf{27.8} & $-2.8$ & --- \\
R65 & $\mathrm{2a} + \mathrm{1ha} \,{\to}\, \mathrm{HCOOH} + \mathrm{(CH_2OH)_2}$ & $\mathrm{HCOOH}$ & 0 & 26.4 & 3/3 & 22.3 & $+3.1$ & 0 & $22.5 \pm 0.3$ & \textbf{22.3} & $-4.1$ & \textbf{24.0} & $-2.4$ & --- \\
R66 & $\mathrm{2a} + \mathrm{1ha} \,{\to}\, \mathrm{HCOOH} + \mathrm{(CH_2OH)_2}$ & $\mathrm{NH_3}$ & 0 & 30.1 & 3/3 & 24.7 & $-8.3$ & 2 & $22.1 \pm 2.2$ & \textbf{20.4} & $-9.7$ & \textbf{20.3} & $-9.8$ & --- \\
R67 & $\mathrm{3a} \,{\to}\, \mathrm{3e}$ & None & 3 & 24.9 & 3/3 & 23.2 & $+6.6$ & 4 & $24.3 \pm 1.9$ & \textbf{22.1} & $-2.7$ & \textbf{23.5} & $-1.3$ & Fig.~\ref{fig:si_plate5} \\
R68 & $\mathrm{3a} \,{\to}\, \mathrm{3e}$ & $\mathrm{HCOOH}$ & 1 & 26.9 & 3/3 & 23.2 & $-1.8$ & 2 & $24.6 \pm 5.7$ & \textbf{18.0} & $-8.9$ & \textbf{19.1} & $-7.8$ & --- \\
R69 & $\mathrm{3a} \,{\to}\, \mathrm{3e}$ & $\mathrm{NH_3}$ & 2 & 17.1 & 3/3 & --- & --- & 1 & $18.7 \pm 1.2$ & \textbf{17.3} & $+0.2$ & \textbf{17.1} & $0.0$ & --- \\
R70 & $\mathrm{3k} \,{\to}\, \mathrm{3e}$ & None & 3 & 27.3 & 3/3 & 26.9 & $-6.4$ & 4 & $26.0 \pm 2.3$ & \textbf{23.4} & $-3.8$ & \textbf{25.3} & $-2.0$ & Fig.~\ref{fig:si_plate5} \\
R71 & $\mathrm{3k} \,{\to}\, \mathrm{3e}$ & $\mathrm{HCOOH}$ & 1 & 29.9 & 3/3 & 20.6 & $-3.5$ & 1 & $23.3 \pm 2.7$ & \textbf{20.6} & $-9.3$ & \textbf{21.9} & $-7.9$ & --- \\
R72 & $\mathrm{3k} \,{\to}\, \mathrm{3e}$ & $\mathrm{NH_3}$ & 2 & 20.2 & 3/3 & 21.8 & $+0.8$ & 3 & $22.5 \pm 1.9$ & \textbf{21.0} & $+0.8$ & \textbf{20.5} & $+0.3$ & --- \\
R73 & $\mathrm{3a} \,{\to}\, \mathrm{3k}$ & None & 1 & 30.0 & 3/3 & 29.9 & --- & 4 & $28.3 \pm 1.7$ & \textbf{26.4} & $-3.6$ & \textbf{27.1} & $-3.0$ & --- \\
R74 & $\mathrm{3e} + \mathrm{CH_2O} \,{\to}\, \mathrm{4k}$ & None & 2 & 17.6 & 3/3 & 13.3 & $-4.9$ & 3 & $14.1 \pm 2.5$ & \textbf{12.1} & $-5.6$ & \textbf{14.9} & $-2.8$ & Fig.~\ref{fig:si_plate5} \\
R75 & $\mathrm{3e} + \mathrm{CH_2O} \,{\to}\, \mathrm{4k}$ & $\mathrm{HCOOH}$ & 0 & 16.6 & 3/3 & 12.8 & $-3.7$ & 1 & $12.8 \pm 2.7$ & \textbf{11.2} & $-5.5$ & \textbf{13.2} & $-3.5$ & --- \\
R76 & $\mathrm{3e} + \mathrm{CH_2O} \,{\to}\, \mathrm{4k}$ & $\mathrm{NH_3}$ & 0 & 13.1 & 3/3 & 10.4 & $0.0$ & 1 & $7.9 \pm 0.1$ & \textbf{7.8} & $-5.3$ & \textbf{10.0} & $-3.1$ & --- \\
R77 & $\mathrm{3e} + \mathrm{CH_2O} \,{\to}\, \mathrm{DHMP}$ & None & 1 & 23.6 & 3/3 & 22.6 & $+2.8$ & 3 & $16.8 \pm 1.9$ & \textbf{14.6} & $-9.0$ & \textbf{17.3} & $-6.3$ & --- \\
R78 & $\mathrm{3e} + \mathrm{CH_2O} \,{\to}\, \mathrm{DHMP}$ & $\mathrm{HCOOH}$ & 0 & 19.5 & 3/3 & 12.5 & $-4.2$ & 2 & $12.4 \pm 0.2$ & \textbf{12.2} & $-7.2$ & \textbf{14.5} & $-5.0$ & Fig.~\ref{fig:si_plate6} \\
R79 & $\mathrm{3e} + \mathrm{CH_2O} \,{\to}\, \mathrm{DHMP}$ & $\mathrm{NH_3}$ & 0 & 15.5 & 3/3 & 14.7 & $+7.6$ & 1 & $9.9 \pm 0.7$ & \textbf{9.4} & $-6.1$ & \textbf{10.7} & $-4.8$ & --- \\
R80 & $\mathrm{4k} \,{\to}\, \mathrm{4e}$ & None & 3 & 29.4 & 2/3 & 23.0 & $-5.5$ & 3 & $23.0 \pm 0.1$ & \textbf{23.0} & $-6.5$ & \textbf{25.9} & $-3.6$ & Fig.~\ref{fig:si_plate5} \\
R81 & $\mathrm{4k} \,{\to}\, \mathrm{4e}$ & $\mathrm{HCOOH}$ & 1 & 28.6 & 3/3 & 20.7 & $-4.0$ & 1 & $22.7 \pm 1.8$ & \textbf{20.7} & $-8.0$ & \textbf{23.4} & $-5.2$ & --- \\
R82 & $\mathrm{4k} \,{\to}\, \mathrm{4e}$ & $\mathrm{NH_3}$ & 2 & 18.5 & 3/3 & 24.6 & $+1.3$ & 1 & $22.5 \pm 0.5$ & \textbf{22.1} & $+3.6$ & \textbf{23.2} & $+4.7$ & Fig.~\ref{fig:si_plate7} \\
R83 & $\mathrm{4a3} \,{\to}\, \mathrm{2a} + \mathrm{2e}$ & None & 2 & 31.5 & 3/3 & 26.4 & $-8.2$ & 3 & $24.8 \pm 3.4$ & \textbf{21.0} & $-10.5$ & \textbf{23.6} & $-7.9$ & --- \\
R84 & $\mathrm{4a3} \,{\to}\, \mathrm{2a} + \mathrm{2e}$ & $\mathrm{HCOOH}$ & 0 & 27.1 & 3/3 & 21.3 & $-11.4$ & 1 & $20.7 \pm 0.6$ & \textbf{20.1} & $-7.0$ & \textbf{22.9} & $-4.2$ & Fig.~\ref{fig:si_plate6} \\
R85 & $\mathrm{4a3} \,{\to}\, \mathrm{2a} + \mathrm{2e}$ & $\mathrm{NH_3}$ & 1 & 26.9 & 3/3 & 17.9 & $-8.5$ & 2 & $18.0 \pm 0.7$ & \textbf{17.3} & $-9.6$ & \textbf{17.6} & $-9.3$ & --- \\
R86 & $\mathrm{4a} \,{\to}\, \mathrm{2a} + \mathrm{2e}$ & None & 2 & 33.6 & 2/3 & --- & --- & 1 & $29.3 \pm 1.1$ & \textbf{28.6} & $-5.0$ & \textbf{31.1} & $-2.5$ & --- \\
R87 & $\mathrm{4a} \,{\to}\, \mathrm{2a} + \mathrm{2e}$ & $\mathrm{HCOOH}$ & 0 & 30.1 & 3/3 & 22.2 & $-9.8$ & 1 & $21.9 \pm 0.5$ & \textbf{21.4} & $-8.7$ & \textbf{21.3} & $-8.7$ & Fig.~\ref{fig:si_plate6} \\
R88 & $\mathrm{4a} \,{\to}\, \mathrm{2a} + \mathrm{2e}$ & $\mathrm{NH_3}$ & 1 & 27.3 & 3/3 & 17.0 & $-9.0$ & 1 & $17.1 \pm 0.1$ & \textbf{17.0} & $-10.3$ & \textbf{19.0} & $-8.3$ & --- \\
R89 & $\mathrm{CH_2O} + \mathrm{NH_3} \,{\to}\, \mathrm{CH_2(OH)NH_2}$ & None & 2 & 7.0 & 3/3 & 0.1 & $+0.6$ & 3 & $-0.3 \pm 0.6$ & \textbf{-0.9} & $-7.9$ & \textbf{1.5} & $-5.5$ & --- \\
R90 & $\mathrm{CH_2O} + \mathrm{H_2S} \,{\to}\, \mathrm{CH_2(OH)SH}$ & None & 2 & 12.2 & 3/3 & 11.9 & $-0.9$ & 3 & $10.8 \pm 0.1$ & \textbf{10.7} & $-1.5$ & \textbf{7.2} & $-4.9$ & --- \\
R91 & $\mathrm{CH_3OH} + \mathrm{NH_3} \,{\to}\, \mathrm{CH_3NH_2} + \mathrm{H_2O}$ & None & 3 & 32.2 & 3/3 & 33.1 & $+0.9$ & 5 & $32.9 \pm 0.2$ & \textbf{32.7} & $+0.5$ & \textbf{30.5} & $-1.7$ & Fig.~\ref{fig:si_plate7} \\
R92 & $\mathrm{CH_3OH} + \mathrm{H_2S} \,{\to}\, \mathrm{CH_3SH} + \mathrm{H_2O}$ & None & 3 & 20.6 & 3/3 & 24.8 & $+0.6$ & 3 & $25.1 \pm 0.5$ & \textbf{24.8} & $+4.2$ & \textbf{17.6} & $-3.1$ & Fig.~\ref{fig:si_plate7} \\
R93 & $\mathrm{HCOOH} + \mathrm{NH_3} \,{\to}\, \mathrm{HCONH_2} + \mathrm{H_2O}$ & None & 2 & 24.1 & 3/3 & 21.1 & $0.0$ & 2-st & $21.1 \pm 0.0$ & \textbf{21.1} & $-3.0$ & \textbf{23.1} & $-1.0$ & --- \\
R94 & $\mathrm{HCOOH} + \mathrm{H_2S} \,{\to}\, \mathrm{HCOSH} + \mathrm{H_2O}$ & None & 2 & 33.2 & 3/3 & 29.9 & $+1.2$ & 2-st & $28.2 \pm 1.5$ & \textbf{27.3} & $-5.8$ & \textbf{27.4} & $-5.8$ & --- \\
R95 & $\mathrm{HCOOH} + \mathrm{CH_3OH} \,{\to}\, \mathrm{HCOOCH_3} + \mathrm{H_2O}$ & None & 2 & 26.9 & 3/3 & 23.3 & $-1.2$ & 2-st & $22.9 \pm 0.5$ & \textbf{22.3} & $-4.6$ & \textbf{22.6} & $-4.4$ & --- \\
R96 & $\mathrm{CH_2O} + \mathrm{CH_2(OH)NH_2} \,{\to}\, \mathrm{HCONH_2} + \mathrm{CH_3OH}$ & None & 0 & 23.3 & 3/3 & 22.2 & --- & 2 & $19.9 \pm 1.8$ & \textbf{17.8} & $-5.5$ & \textbf{18.2} & $-5.1$ & --- \\
R97 & $\mathrm{CH_2O} + \mathrm{CH_2(OH)SH} \,{\to}\, \mathrm{HCOSH} + \mathrm{CH_3OH}$ & None & 0 & 28.9 & 3/3 & 28.2 & --- & 2 & $27.7 \pm 0.2$ & \textbf{27.5} & $-1.4$ & \textbf{27.9} & $-1.0$ & --- \\
\end{longtable}
}

\end{document}